\documentclass[fleqn,usenatbib]{mnras}

\usepackage{newtxtext,newtxmath}
\usepackage{threeparttable}
\usepackage{multirow}
\usepackage{xurl}
\usepackage{booktabs}

\usepackage[T1]{fontenc}
\usepackage{soul}
\DeclareRobustCommand{\VAN}[3]{#2}
\let\VANthebibliography\thebibliography
\def\thebibliography{\DeclareRobustCommand{\VAN}[3]{##3}\VANthebibliography}

\usepackage{graphicx}	
\usepackage{amsmath}	

\makeatletter
\newcommand*{\linktocite}[2]{%
  \hyper@natlinkstart{#1}#2\hyper@natlinkend}
\makeatother

\usepackage{orcidlink}

\title[HAT-P-18\,b with JWST NIRISS/SOSS]{Near-Infrared Transmission Spectroscopy of HAT-P-18\,b with NIRISS: Disentangling Planetary and Stellar Features in the Era of JWST}
\author[Marylou Fournier-Tondreau et al.]{Marylou Fournier-Tondreau\orcidlink{0000-0002-5428-0453}$^{1}$\thanks{E-mail: marylou.fournier.tondreau@umontreal.ca}, Ryan J. MacDonald\orcidlink{0000-0003-4816-3469}$^{2}$\thanks{NHFP Sagan Fellow}, Michael Radica\orcidlink{0000-0002-3328-1203}$^{1}$, David Lafrenière\orcidlink{0000-0002-6780-4252}$^{1}$, Luis
\newauthor{Welbanks\orcidlink{0000-0003-0156-4564}$^{3}$\thanks{NHFP Sagan Fellow}, Caroline Piaulet\orcidlink{0000-0002-2875-917X}$^{1}$, Louis-Philippe Coulombe\orcidlink{0000-0002-2195-735X}$^{1}$, Romain Allart\orcidlink{0000-0002-1199-9759}$^{1}$\thanks{Trottier Postdoctoral Fellow}, Kim Morel\orcidlink{0000-0002-1901-1266}$^{1}$,} \newauthor{Étienne Artigau\orcidlink{0000-0003-3506-5667}$^{1,4}$, Loïc Albert\orcidlink{0000-0003-0475-9375}$^{1}$, Olivia Lim\orcidlink{0000-0003-4676-0622}$^{1}$, René Doyon\orcidlink{0000-0001-5485-4675}$^{1}$, Björn Benneke\orcidlink{0000-0001-5578-1498}$^{1}$, Jason F. Rowe\orcidlink{0000-0002-5904-1865}$^{5}$,} \newauthor{Antoine Darveau-Bernier\orcidlink{0000-0002-7786-0661}$^{1}$, Nicolas B. Cowan\orcidlink{0000-0001-6129-5699}$^{6,7}$, Nikole K. Lewis\orcidlink{0000-0002-8507-1304}$^{8}$, Neil James Cook\orcidlink{0000-0003-4166-4121}$^{1}$, Laura} \newauthor{Flagg\orcidlink{0000-0001-6362-0571}$^{8}$, Frédéric Genest\orcidlink{0000-0003-0602-9106}$^{1}$, Stefan Pelletier\orcidlink{0000-0002-8573-805X}$^{1}$, Doug Johnstone\orcidlink{0000-0002-6773-459X}$^{9,10}$, Lisa Dang\orcidlink{0000-0003-4987-6591}$^{1}$\thanks{Banting Postdoctoral Fellow}, Lisa Kaltenegger\orcidlink{0000-0002-0436-1802}$^{8}$,}\newauthor{Jake Taylor\orcidlink{0000-0003-4844-9838}$^{1,11}$, Jake D. Turner\orcidlink{0000-0001-7836-1787}$^{8}$\thanks{NHFP Sagan Fellow}}\\
$^{1}$Institut Trottier de recherche sur les exoplanètes and Département de Physique, Université de Montréal, 1375 Avenue Thérèse-Lavoie-Roux,\\ Montréal, QC, H2V 0B3, Canada\\
$^{2}$Department of Astronomy, University of Michigan, 1085 S. University Ave., Ann Arbor, MI, 48109, USA\\
$^{3}$School of Earth and Space Exploration, Arizona State University, 781 Terrace Mall, Tempe, AZ, 85287, USA\\
$^{4}$Observatoire du Mont-Mégantic, Université de Montréal, Montréal, QC, H3C 3J7, Canada\\
$^{5}$Department of Physics \& Astronomy, Bishop’s University, Sherbrooke, QC, J1M 1Z7, Canada\\
$^{6}$Department of Physics, McGill University, 3600 rue University, Montréal, QC, H3A 2T8, Canada\\
$^{7}$Department of Earth and Planetary Sciences, McGill University, 3600 rue University, Montréal, QC, H3A 2T8, Canada\\
$^{8}$Carl Sagan Institute and Department of Astronomy, Cornell University, Ithaca, NY, 14853, USA\\
$^{9}$Department of Physics and Astronomy, University of Victoria, Victoria, BC, V8P 5C2, Canada\\
$^{10}$NRC Herzberg Astronomy and Astrophysics, 5071 West Saanich Rd, Victoria, BC, V9E 2E7, Canada\\
$^{11}$Department of Physics, University of Oxford, Parks Rd, Oxford, OX1 3PU, UK\\
}
\date{Accepted 2023 November 26. Received 2023 November 01; in original form 2023 July 26}

\pubyear{2023}

\begin{document}
\label{firstpage}
\pagerange{\pageref{firstpage}--\pageref{lastpage}}

\maketitle

\begin{abstract}
The JWST Early Release Observations (ERO) included a NIRISS/SOSS (0.6--2.8\,$\mu$m) transit of the $\sim$\,850\,K Saturn-mass exoplanet HAT-P-18\,b. Initial analysis of these data reported detections of water, escaping helium, and haze. However, active K dwarfs like HAT-P-18 possess surface heterogeneities --- starspots and faculae --- that can complicate the interpretation of transmission spectra, and indeed, a spot-crossing event is present in HAT-P-18\,b's NIRISS/SOSS light curves. Here, we present an extensive reanalysis and interpretation of the JWST ERO transmission spectrum of HAT-P-18\,b, as well as HST/WFC3 and \textit{Spitzer}/IRAC transit observations. We detect H$_2$O (12.5\,$\sigma$), CO$_2$ (7.3\,$\sigma$), a cloud deck (7.4\,$\sigma$), and unocculted starspots (5.8\,$\sigma$), alongside hints of Na (2.7\,$\sigma$). We do not detect the previously reported CH$_4$ ($\log$ CH$_4$ $<$ -6 to 2\,$\sigma$). We obtain excellent agreement between three independent retrieval codes, which find a sub-solar H$_2$O abundance ($\log$ H$_2$O $\approx -4.4 \pm 0.3$). However, the inferred CO$_2$ abundance ($\log$ CO$_2$ $\approx -4.8 \pm 0.4$) is significantly super-solar and requires further investigation into its origin. We also introduce new stellar heterogeneity considerations by fitting for the active regions' surface gravities --- a proxy for the effects of magnetic pressure. Finally, we compare our JWST inferences to those from HST/WFC3 and \textit{Spitzer}/IRAC. Our results highlight the exceptional promise of simultaneous planetary atmosphere and stellar heterogeneity constraints in the era of JWST and demonstrate that JWST transmission spectra may warrant more complex treatments of the transit light source effect. 
\end{abstract}




\section{Introduction}
\label{sec: intro}
In the works for over two decades, the James Webb Space Telescope (JWST) is finally operational. Astronomers can now count on space instruments with modes designed to study exoplanetary atmospheres. Transmission spectroscopy is a commonly used method to reveal the composition and structure of an atmosphere and, therefore, to enable inferences about a planet's formation and evolution history. During a transit, part of the starlight is blocked by the planet, whereas some light passes through the planetary atmosphere and can introduce measurable absorption features \citep{seager2000,brown2001b}. The resulting transmission spectrum can reveal key information regarding the abundance of molecular and atomic species and the presence of clouds and hazes \citep[e.g.,][]{charbonneau2002,tinetti2007,wakeford2015,sing2016}. 

Astronomers have faced many challenges when conducting atmospheric studies with the Hubble (HST) and Spitzer Space Telescopes since neither observatory was designed for exoplanet observations. Numerous technical difficulties, instrument systematics, as well as narrow wavelength range and spectral resolution, have limited atmospheric inferences. For example, observations with the Wide Field Camera 3 (WFC3) instrument aboard HST are complicated by systematic trends, such as ``HST breathing'' effects, visit-long slopes, and the ``ramp'' effect \citep{wakeford2016}. Nevertheless, the efforts and ingenuity of scientists have led to astonishing discoveries, such as the detection of several chemical species, including water vapour on hot Jupiters (e.g., \citealp{tinetti2007,tsiaras2018}) and even on a sub-Neptune \citep[e.g.,][]{benneke2019b}, the inference of clouds and hazes in several gas giants (e.g., \citealp{wakeford2015,sing2016}), and the observation of atmospheric escape on hot Neptune-like exoplanets (e.g., \citealp{ehrenreich2015,spake2018}).

HAT-P-18\,b was discovered in 2010 by \citet{hartman2011} using the Hungarian-made Automated Telescope Network. It is of approximately Saturn mass ($M = 0.197\,M\textsubscript{J}$), but with an inflated radius ($R = 0.995\,R\textsubscript{J}$), due to its higher temperature ($T\textsubscript{eq} = 852$\,K) relative to the Solar System giants, which is a consequence of the planet's comparatively short orbital period ($P = 5.5$\,days). Ground- and space-based transmission spectroscopy has been performed on this target. \citet{kirk2017} suggested a high-altitude haze consistent with the detection of Rayleigh scattering and the absence of the sodium absorption feature using the Auxiliary-port CAMera (ACAM) instrument on the William Herschel Telescope (WHT). \citet{tsiaras2018} detected the presence of water vapour and a grey, opaque cloud deck in HAT-P-18\,b's atmosphere using HST/WFC3, reporting a water abundance of
$\log{\rm{H_2O}} = -2.63 \pm 1.18$ and a cloud-top pressure of log $P$\textsubscript{cloud} [Pa] = $2.82 \pm 0.91$. \citet{fu2022} presented an analysis of the transit observed in the Single Object Slitless Spectroscopy (SOSS) mode \citep{albert2023} of the Near Infrared Imager and Slitless Spectrometer (NIRISS) instrument \citep{doyon2023} on board the JWST and detected water (with an abundance of $\log{\rm{H_2O}} = -3.03_{-0.25}^{+0.31}$), hints of methane ($\Delta\log Z$ = 3.79, or a 3.2\,$\sigma$ confidence), as well as excess helium absorption and tail in an otherwise very hazy atmosphere.

HAT-P-18 is an active K dwarf with an effective temperature of 4803~K and a slightly super-solar metallicity ([Fe/H] = 0.10 $\pm$ 0.08; \citealp{hartman2011}). Stellar active regions, such as starspots and faculae, can introduce spectral features in transmission spectra that overlap those of exoplanetary atmospheres. Occulted active regions were often masked when fitting transit models to spectroscopic light curves; however, the impact of the occulted spot on the transmission spectrum is still present despite that \citep[e.g.,][]{oshagh2014,bixel2019}. This can also lead to a biased transmission spectrum by impacting not only the transit depth but possibly the mid-transit time, the scaled semi-major axis, and the impact parameter \citep[e.g.,][]{barros2013,alexoudi2020}. Recent studies have moved towards joint inferences of transit and active region properties (e.g., \citealp{bixel2019,espinoza2019a}). The NASA Study Analysis Group on the effect of stellar contamination on space-based transmission spectroscopy (SAG 21) of the Exoplanet Exploration Program Analysis Group (ExoPAG) recommends performing these joint inferences with future observations instead of masking active region occultations \citep{rackham2023a}.  

Unocculted stellar active regions have long been recognized as a significant obstacle to exoplanet transmission spectroscopy. Early HST studies recognized that unocculted cool starspots can cause strong transit depth slopes towards short visible wavelengths \citep[e.g.,][]{pont2007,pont2013,mccullough2014}. Similarly, unocculted hot faculae can imprint a negative slope in transmission spectra \citep[e.g.,][]{rackham2018,kirk2021}. The physical origin of this `stellar contamination' is a mismatch between the intensity of the stellar surface sampled by the planet during transit and the average spectrum of the star (including the photosphere, starspots, and faculae). Since the contrast ratio between the spectra of stellar regions at different temperatures increases as shorter wavelengths, this `transit light source effect' \citep[TLSE;][]{rackham2018,barclay2021} is wavelength-dependent and more significant at visible wavelengths. The most common approach to deal with the TLSE in early studies was to correct the transmission spectrum based on activity monitoring or occulted starspot properties \citep[e.g.,][]{pont2008,berta2011,sing2011}. \citet{barstow2015} demonstrated that not accounting for starspots when modelling transmission spectra of giant planets can bias retrieved molecular abundances, while \citep{rackham2018} further showed that the TLSE can dominate over absorption features for terrestrial planets. \linktocite{moran2023}{Moran \& Stevenson et al.} \citeyear{moran2023} provide a recent example of this prediction, finding degenerate interpretations between unocculted starspots and atmospheric H$_2$O for JWST observation of a terrestrial exoplanet.

Recent years have seen a renewed focus on incorporating unocculted stellar regions into atmospheric retrieval codes, allowing \emph{simultaneous} inferences of stellar and atmospheric properties. \citet{pinhas2018} developed a transmission spectrum retrieval framework to jointly fit a single population of unocculted stellar heterogeneities and a planetary atmosphere. Subsequent retrieval studies have explored the fidelity of TLSE retrieval assumptions \citep[e.g.,][]{iyer2020,thompson2023,rackham2023b} and applied these joint retrievals to interpret observations from HST and \textit{Spitzer} \citep[e.g.,][]{bruno2020,rathcke2021}, ground-based telescopes \citep[e.g.,][]{bixel2019,jiang2021,kirk2021}, and JWST \linktocite{moran2023}{(Moran \& Stevenson et al.} \citeyear{moran2023}). However, the SAG 21 report \citep{rackham2023a} notes that there is considerable scope to improve the realism and complexity of retrieval prescriptions for unocculted stellar active regions.

In this work, we aim to disentangle stellar and planetary atmosphere signals by including stellar heterogeneities in transit fits and atmospheric retrievals. We present and compare two independent atmospheric reanalyses of HAT-P-18\,b, one using JWST NIRISS/SOSS Early Release Observations (ERO) transit observation and another combining transit observations from HST/WFC3 and the Infrared Array Camera (IRAC) of \textit{Spitzer}. We describe the observations and data reduction approach in Section~\ref{sec:data} and detail our JWST NIRISS/SOSS light curve fitting and occulted starspot analysis in Section~\ref{sec:lightcurve}. Section~\ref{sec:retrievals} describes our joint stellar heterogeneity and atmospheric retrieval method and presents results from three independent retrieval codes. We summarize and discuss our results in Section~\ref{sec:discussion} and conclude in Section~\ref{sec:conclu}.

\section{Observations \& Data Reduction}\label{sec:data}
The scientific legacy of HST and \textit{Spitzer} is considerable, particularly for exoplanet studies; JWST will, in many ways, build on this legacy. In this work, we present transit observations with JWST and its predecessors, HST and \textit{Spitzer}, to show the potential of NIRISS/SOSS to characterize exoplanetary atmospheres and to cope with challenges such as stellar contamination.

\subsection{Observations}
\begin{table}
\begin{center}
\caption{\label{tab1} Parameters of the HAT-P-18 planetary system used in this analysis}
\begin{tabular}{l|cc}\toprule
Parameters & HAT-P-18 & Units\\ \midrule
\multicolumn{3}{l}{\textbf{Stellar parameters}}\\\midrule
Spectral type  & K2V & \\
Stellar radius & 0.749 $\pm$ 0.037 & R$_\odot$ \\
Stellar mass & 0.770 $\pm$ 0.031 & M$_\odot$\\
Metallicity & 0.10 $\pm$ 0.08 & [Fe/H] \\
Stellar surface gravity & 4.57 $\pm$ 0.04 & log$_{10}$ cm/s$^2$\\
Effective temperature & 4803 $\pm$ 80& K\\ \midrule
\multicolumn{3}{l}{\textbf{Planetary and transit parameters}}\\ \midrule
Planet radius  & 0.995 $\pm$ 0.052 & R$_{\text{J}}$\\
Planet mass  &0.197 $\pm$ 0.013&M$_{\text{J}}$\\
Orbital period  &5.508023 $\pm$ 0.000006 & day\\
Eccentricity  & 0.084 $\pm$ 0.048 & \\
Argument of periastron & 120.0 $\pm$ 56.0 & $\deg$\\
Impact parameter &0.324$^{+0.055}_{-0.078}$&\\
Scaled semi-major axis & 16.04 $\pm$ 0.75 & \\
Transit duration& 0.1131 $\pm$ 0.0009 & day \\
Scaled planet radius &0.1365 $\pm$ 0.0015 &\\
Equilibrium temperature  & 852 $\pm$ 28 & K\\\bottomrule
\\
\multicolumn{3}{l}{\footnotesize \textit{Note:} Parameters from \citealp{hartman2011}} \\
\end{tabular}
\end{center}
\end{table}
\subsubsection{JWST NIRISS/SOSS}

HAT-P-18\,b was observed in transit using the SOSS mode of the NIRISS instrument \citep{albert2023,doyon2023} on board JWST as part of the ERO program (PI: Klaus M. Pontoppidan; \citealp{pontoppidan2022}). The time series observation (TSO) started on June 13\textsuperscript{th}, 2022, at 04:36:50.861 UTC and lasted 7.15\,hours, which covered the 2.7\,hours transit as well as some baseline before and after the transit. The GR700XD grism and the CLEAR filter were used, along with the SUBSTRIP256 subarray, which captures both diffraction orders 1 and 2. There are 469 integrations, each consisting of 9 groups and lasting 54.94\,seconds. In addition, a second exposure with the GR700XD grism and F277W filter was taken directly after the end of the main CLEAR filter TSO. This exposure had the exact same readout parameters as the CLEAR exposure, except it consisted of only 11 integrations lasting 10.07 minutes. The TSO was previously published by \citet{fu2022}, the major findings of which are summarized in Section~\ref{sec: intro}.

\subsubsection{HST/WFC3 + Spitzer/IRAC}

Transit observations of HAT-P-18\,b were obtained using HST/WFC3 with the G141 grism in the spatial scan mode. These come from two visits, consisting of five orbits each, made as part of the Cycle 23 HST General Observer campaign (PI: Drake Deming; \citealt{deming2015}) in February 2016 and January 2017. These observations spanned the wavelength range from 1.1 to 1.7\,$\mu$m and were downloaded from the Mikulski Archive for Space Telescopes (MAST). The raw light curves show characteristic, non-linear systematics because of thermal effects induced by the telescope's 96-minute orbit \citep{foster1995}. This leads us to follow standard practice (e.g., \citealp{benneke2019b}) and discard from our analysis the first orbit as well as the first two measurements in each subsequent orbit. This leaves one orbit on either side of the transit and two orbits corresponding to the transit itself, including significant portions of the ingress and egress. These HST transits were previously published by \citet{tsiaras2018}.

Transit observations with \textit{Spitzer} come from two visits in 2013 using the IRAC instrument in the 3.6 and 4.5\,$\mu$m channels (PI: Jean-Michel Désert; \citealt{desert2012}). The data were obtained from the Spitzer Heritage Archive.

\subsection{Data Reduction}

\subsubsection{JWST NIRISS/SOSS Data Reduction}\label{sec:reduction_jwst}

We reduced the NIRISS/SOSS TSO using the \texttt{supreme-SPOON} pipeline \citep{feinstein2022, coulombe2022,radica2023,lim2023}, starting from the raw, uncalibrated files downloaded from the MAST archives. Some steps in the \texttt{supreme-SPOON} pipeline are handled by the official JWST pipeline. We follow a nearly identical procedure for the reduction as was presented in \citet{feinstein2022} and \citet{radica2023}: we first process the TSO through \texttt{supreme-SPOON} Stage 1, which performs the detector level calibrations. This includes, in particular, the subtraction of the column-correlated 1/$f$ noise, during which we ensure to properly mask several undispersed (zeroth order) as well as the sole dispersed (likely a second order, slightly above the target third order) contaminants of background field stars, and the ramp fitting to convert from 3D ``ramp'' to 2D ``slope'' images.

\begin{figure*}
	\centering
	\includegraphics[width=\textwidth]{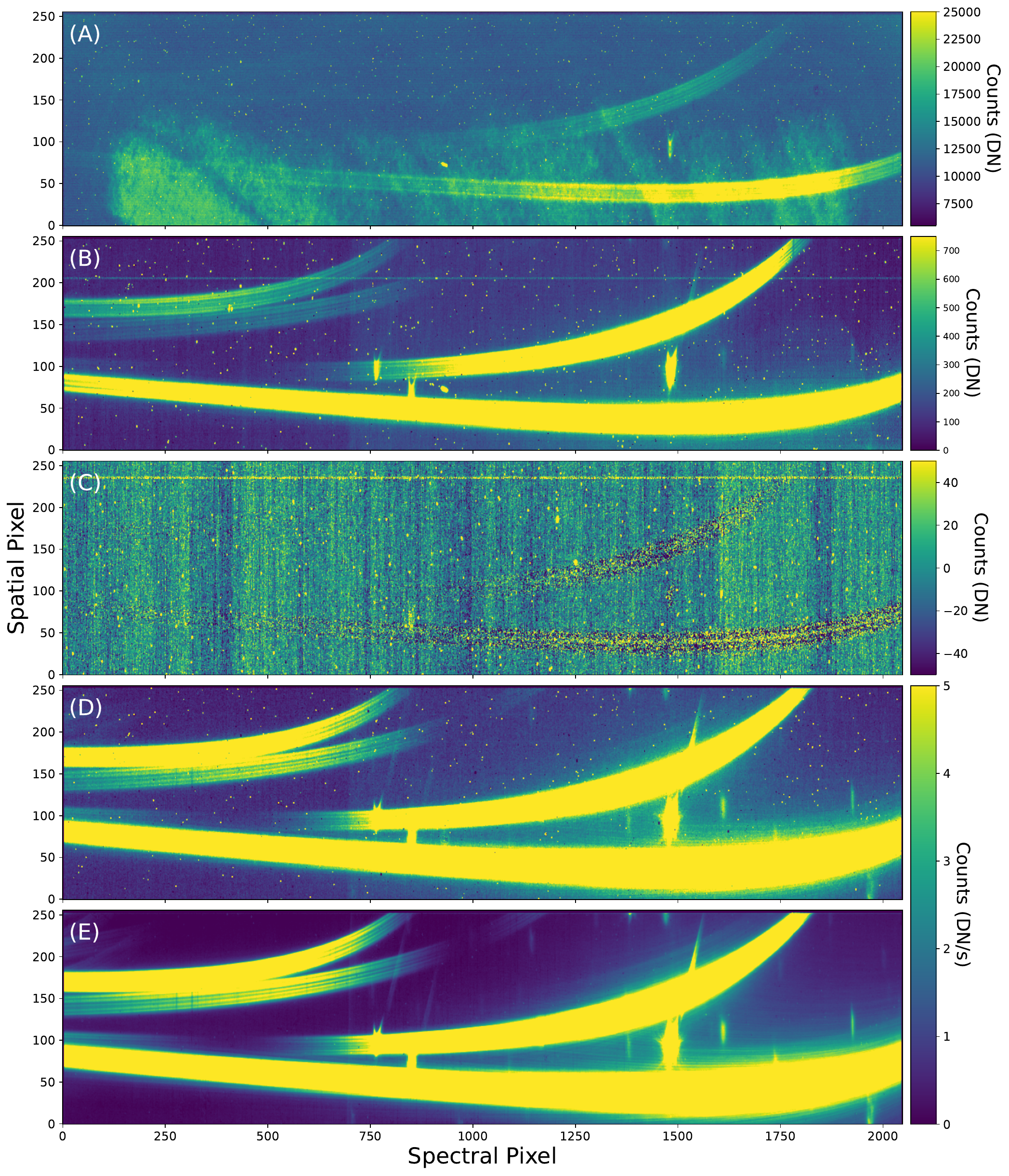}
    \caption{Data products at different stages of the reduction process.
    \textbf{(A)}: A raw, uncalibrated data frame in data numbers (DN).
    \textbf{(B)}: Data frame after superbias subtraction and reference pixel correction. 
    \textbf{(C)}: 1/$f$ noise.
    \textbf{(D)}: After ramp fitting and flat field correction. 
    \textbf{(E)}: Final calibrated data product after background subtraction and bad pixel correction.
    \label{fig:Reduction Steps}}
\end{figure*}

Stage 2 of \texttt{supreme-SPOON} performs further reductions on the slope-level products, including the background subtraction, for which we use the SOSS SUBSTRIP256 background model provided by the Space Telescope Science Institute (STScI)\footnote{\url{https://jwst-docs.stsci.edu/jwst-calibration-pipeline-caveats/jwst-time-series-observations-pipeline-caveats/niriss-time-series-observations-pipeline-caveats\#NIRISSTimeSeriesObservationsPipelineCaveats-SOSSskybackground}}, and tracing of the target orders to define the extraction boxes as well as the stability of the target trace (changes in $x$ and/or $y$ position, changes in morphology, etc.) over the course of the TSO. For further details on these steps, please see \citet{radica2023}. We note, though, that using a constant scaling of the STScI background model, such as in \citet{radica2023}, did not perfectly remove the background step. We, therefore, separately scaled the STScI model blueward and redward of the step \citep[e.g.,][]{lim2023}. We find values for the best-fitting background scaling to be 0.4384 and 0.4080 for pre- and post-jump, respectively. 

We extract the stellar spectra using the \texttt{ATOCA} algorithm \citep{darveau-bernier2022} using a \texttt{specprofile} reference file created specifically for this observation using the \texttt{APPLESOSS} code \citep{radica2022}. \texttt{ATOCA} explicitly takes into account the overlap between the first and second orders of the target spectrum on the detector during the extraction, though the expected dilution introduced from this overlap is predicted to be near-negligible for relative measurements such as exoplanet transmission spectroscopy \citep{darveau-bernier2022,radica2022}. 

We use the updated \texttt{APPLESOSS~v2.0.0} in this work. The initial version of \texttt{APPLESOSS} presented in \citet{radica2022} used the WebbPSF package \citep{perrin2014} to simulate the extended wings of the SOSS PSF (Point Spread Function). However, due to concerns that the simulated PSFs underpredict the SOSS wings\footnote{JWST Technical Report JWST-STScI-008270}, we update the \texttt{APPLESOSS} framework to use the wings of order 1 profiles bluewards of $\sim$\,1.1\,µm where there is no contribution from order 2 on the detector. We note, however, that the transmission spectrum is unchanged whether simulated or empirical wings are used, as the strength of the self-dilution is significantly smaller than the transit depth precision. Finally, after the extraction, we clip any data points which deviate by more than 5\,$\sigma$ from their neighbours in the time direction; $<$\,2\% of points are rejected in this way. We find no significant deviation in the target trace position with the positions included in the \texttt{spectrace} reference file\footnote{\texttt{spectrace} reference file \texttt{jwst\_niriss\_spectrace\_0023.fits} was used in this work.}, and we, therefore, use the default SOSS wavelength solution. A summary of the major reduction steps is shown in Figure \ref{fig:Reduction Steps}.

HAT-P-18 has a co-moving white dwarf companion separated by only $\sim$\,$2.66\arcsec$ at a position angle of $\sim$\,$186\degr$, as revealed by GAIA astrometry \citep{mugrauer2021}\footnote{This companion was also previously reported as a candidate by \citet{ginski2016} based on lucky imaging.}. Given the aperture position angle of the SOSS observation ($252.09\degr$), the spectral trace of this companion is offset by $(-9,+40)$\,pixels relative to the trace of HAT-P-18, i.e., mostly out of the extraction aperture throughout order 1 and for all wavelengths $<0.85\,\mu$m at order 2. Combined with its measured contrast of 8\,mag at $1.4\,\mu$m \citep{mugrauer2021}, it has a negligible effect on the flux extracted for HAT-P-18. No particular action was thus taken to deal with it.

\subsubsection{HST/WFC3 + Spitzer/IRAC Data Reduction \& Light Curve Fitting}

Following standard procedure for HST/WFC3 observations (e.g., \citealp{deming2013,benneke2019b}), we perform the necessary data reduction and light curve fitting by using the modular Exoplanet Transit Eclipses and Phase curves (\texttt{ExoTEP}) framework \citep{benneke2017,benneke2019a}, which employs the \texttt{batman} transit light curve model \citep{kreidberg2015}. Using a Markov chain Monte Carlo (MCMC) method with the \texttt{emcee} package \citep{foreman-mackey2013}, \texttt{ExoTEP} jointly fits the transit and systematics models for the two visits along with photometric noise parameters. We construct spectrophotometric light curves by spectrally binning at 30\,nm intervals \citep{kreidberg2014} the extracted flux for each visit. This binning was chosen because it offered the best trade-off between the number of data versus error bar size, as compared to 10 and 60\,nm bin widths. We follow \citet{benneke2019b} for the procedure of the white and spectrophotometric light curve fits. The latter for both visits are shown in Figure \ref{fig:HST lc}, whereas the best-fitting values from the white light curve fit are quoted in Table \ref{tab: WLC Parameters}. The resulting HST/WFC3 transmission spectrum is displayed in Figure \ref{fig:transmission_spectra}. 

We follow standard procedure for \textit{Spitzer}/IRAC image processing \citep[e.g.,][]{benneke2019b}. As was done for the HST data, we employ \texttt{ExoTEP} to reduce the data further and fit the \textit{Spitzer} light curves. After correcting for the \textit{Spitzer}-specific systematics, we obtain a final light curve for each of the two channels as displayed in Figure \ref{fig:Spitzer lc}. We used 80-second bins for the time axis due to the relatively high cadence of the \textit{Spitzer} data. Compared to the HST spectrophotometric light curves, we find that the scatter is higher, resulting in larger errors for the \textit{Spitzer} data points in the combined transmission spectrum. We kept the transit depths from \textit{Spitzer}/IRAC in the HST transmission spectrum in order to better constrain the atmospheric retrievals.

\section{JWST NIRISS/SOSS Light Curve Fitting \& Occulted Starspot Analysis}\label{sec:lightcurve}

\subsection{Light Curve Fitting with Spot-Crossing Masked}\label{sec:lightcurve_mask}

For the SOSS data, we construct separate white light curves for orders 1 and 2 by summing the flux from all wavelengths in order 1 (0.85--2.8\,µm) and from 0.6--0.85\,µm in order 2. We then fit a transit model to each white light curve using the \texttt{juliet} package \citep{espinoza2019b}. We fix the orbital period to 5.508023\,d, the eccentricity to 0.084, and the argument of periastron to 120$\degr$ (\citealp{hartman2011}; all their values used in this paper are listed in Table \ref{tab1}), and put wide, uninformative priors on the time from mid-transit, $t_0$, the impact parameter, $b$, the scaled orbital semi-major axis, $a/R_*$, and the scaled planet radius, $R_p/R_*$. We fit two parameters of the quadratic limb darkening law following the parameterization of \citet{kipping2013}, a scalar jitter term, $\sigma$, which replaces the flux errors reported by the reduction pipeline, and finally, a term to fix the zero point of the transit baseline. Unlike previous SOSS TSOs \citep{feinstein2022, radica2023}, we find that no detrending is necessary, as the white light curves for both orders are best fit by a transit model with no additional systematics models: $\Delta\log Z$ = 26.1, or a $>$5\,$\sigma$ preference by the \citet{benneke2013} scale. 

\begin{figure*}
    \centering
    \includegraphics[width=0.9\textwidth]{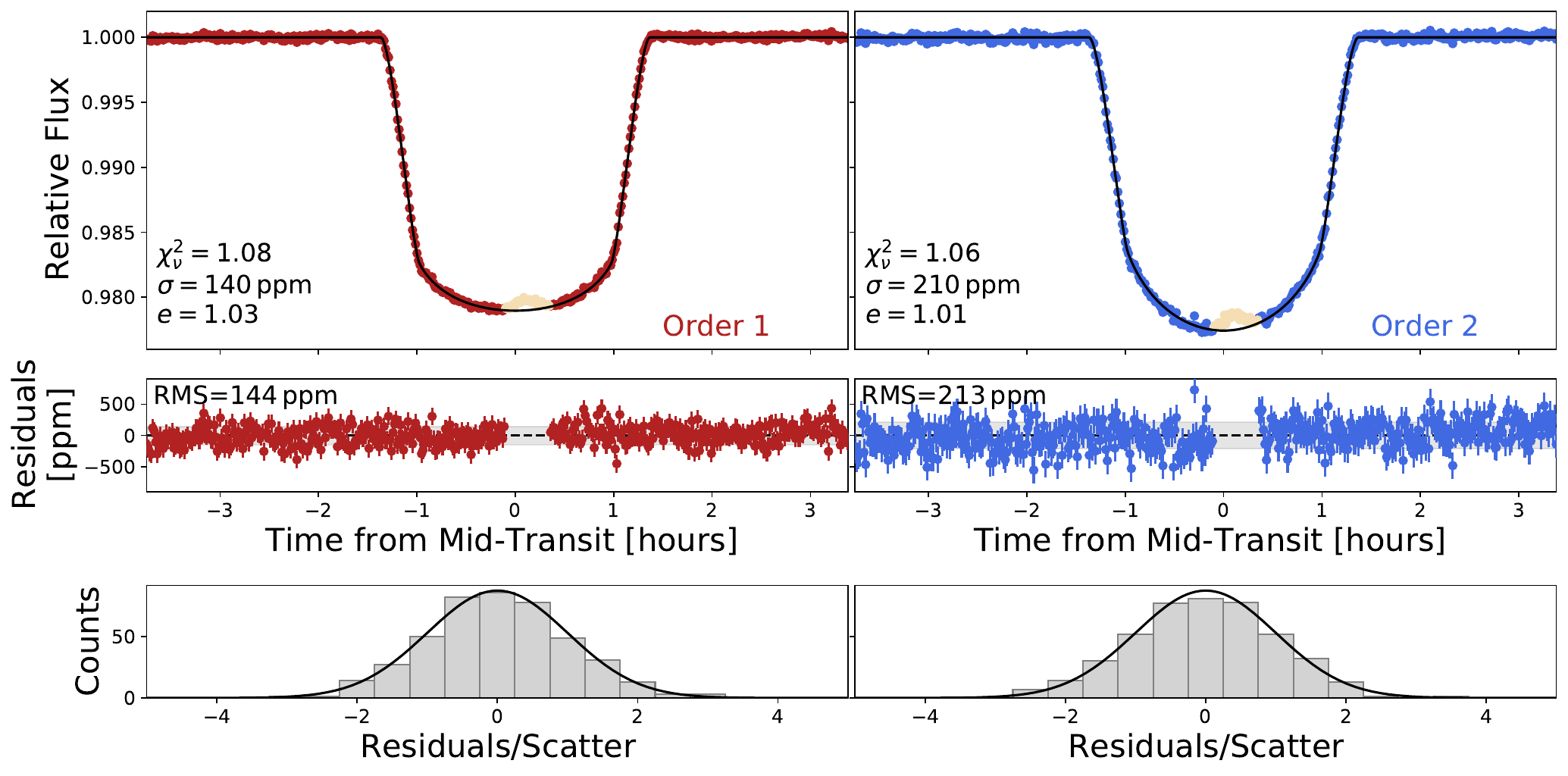}
    \caption{\textit{Top}: NIRISS/SOSS white light curves for order 1 (left) and order 2 (right) with the best-fitting transit model overplotted in black. The integrations associated with the spot-crossing are shown in beige. The fit statistics are indicated for each order: $\chi^2_{\nu}$; the reduced Chi-squared, $\sigma$; the average error bar, and \emph{e}; the error multiple to obtain a $\chi^2_{\nu}$ equal to unity. \emph{Middle}: Residuals to the transit fits. The RMS scatter in the residuals is indicated for each order. \emph{Bottom}: Histogram of residuals.
    \label{fig:WLC}}
\end{figure*}

There is a spot-crossing event clearly visible just after the time from mid-transit. Here we do not include a model of this occulted starspot and instead mask the integrations associated with the spot-crossing (integrations 240 -- 270). Therefore, we fit eight parameters for each order. The reduced Chi-squared statistic for the fits is $\chi^2_\nu$ = 1.08 for order 1 and 1.06 for order 2. The best-fitting transit models for orders 1 and 2 are overplotted in Figure \ref{fig:WLC}.

\begin{figure*}
	\centering
	\includegraphics[width=\textwidth]{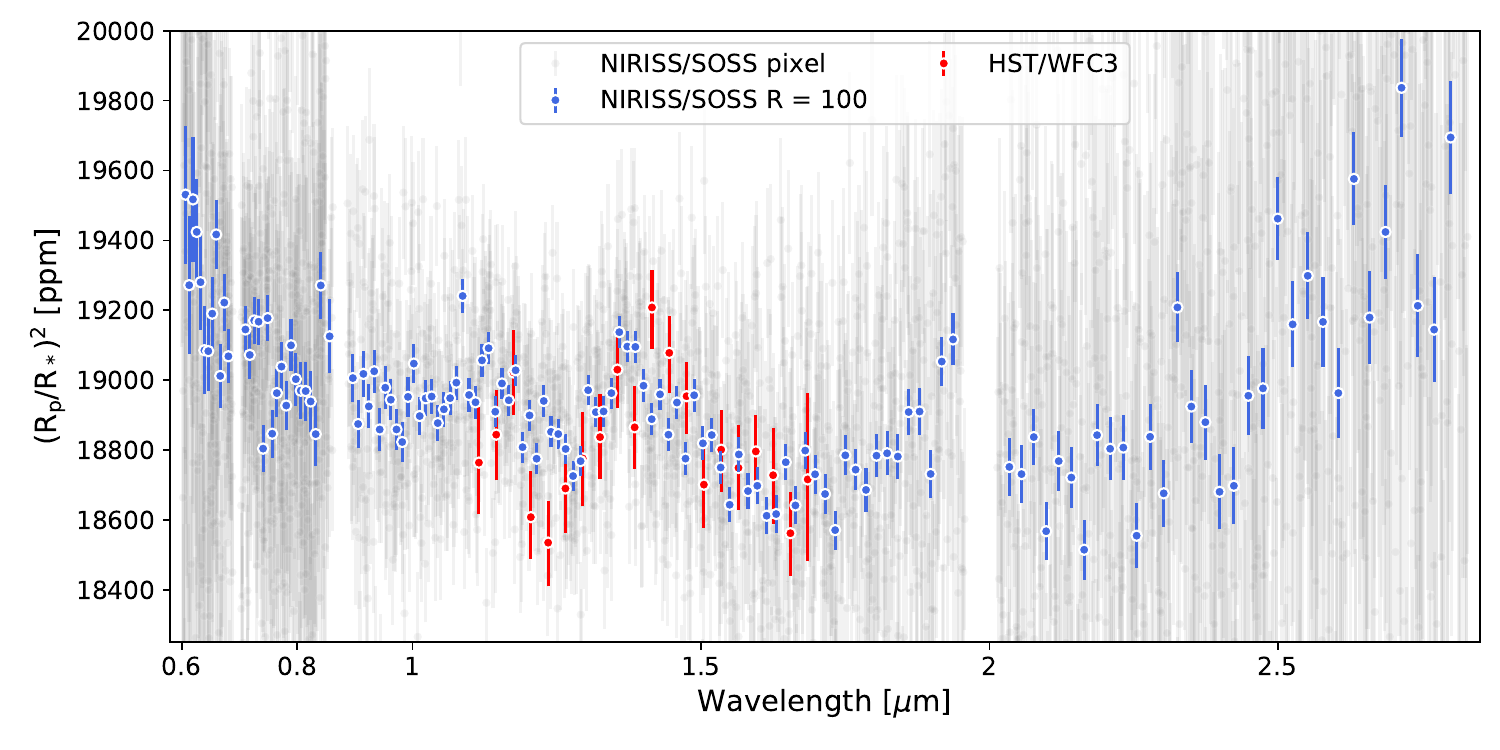}
    \caption{Transmission spectra of HAT-P-18\,b; one with JWST NIRISS/SOSS at pixel resolution (faded grey) and binned to a resolving power of $R$ = 100 (blue) and another with HST/WFC3 (red). Note that no offset was applied to the HST spectrum.}
    \label{fig:transmission_spectra}
\end{figure*}

We then proceed to fit the spectrophotometric light curves at the pixel level --- that is, we fit one light curve per pixel column on the detector. This results in 2038 light curves for order 1 and 567 for order 2. For these fits, we fix the orbital parameters and the transit baseline's zero point to their best-fitting values from the order 1 white light curve fit because of the better signal-to-noise (S/N). We leave only the scaled planet radius, limb-darkening parameters, and jitter term free. We put Gaussian priors on the limb-darkening parameters based on calculations from the \texttt{ExoTiC-LD} package \citep{wakeford2022} using the 3D stagger grid \citep{magic2015}. The widths of the Gaussian priors are set to 0.1 \citep[e.g.,][]{pontoppidan2022,espinoza2022}. We tested larger widths of 0.2 following \citet{patel2022} and the retrieved transmission spectrum is consistent within less than 1\,$\sigma$\footnote{We also tested a retrieval on that transmission spectrum, and all retrieved parameters are consistent within 1\,$\sigma$ to those retrieved with the main transmission spectrum (see Section \ref{sec:retrievals}).}. We also experimented with freely fitting the limb darkening coefficients and found that this results in a slight ($\sim$20\,ppm) offset to the spectrum while preserving the relative amplitudes of the spectral features. We again mask the integrations associated with the spot-crossing in each fit. The resulting transmission spectrum is displayed in Figure \ref{fig:transmission_spectra}.

Although we do not directly use the F277W exposure for science purposes in this study, as in \citet{radica2023}, it is useful to pinpoint undispersed (order 0) contaminants of field stars. Several bright contaminants are clearly visible in the F277W data frame, many of which are also readily visible in the CLEAR frame. However, unlike in \citet{radica2023}, we are unable to post-process our transmission spectrum to correct for the diluting effects of field star contamination. The contaminants are too bright preventing good reconstruction of the target trace. Moreover, as pointed out by \citet{fu2022}, one contaminant is time-varying. We, therefore, mask the affected regions of the transmission spectrum as was done in \citet{feinstein2022} and \citet{fu2022}. The wavelength regions masked in this way are as follows: 0.714 -- 0.724\,µm and 0.841 -- 0.85\,µm in order 2, and 0.853 -- 0.870\,µm, 1.048 -- 1.061\,µm, 1.366 -- 1.384\,µm, and 1.972 -- 2.011\,µm in order 1. A frame of the F277W exposure, highlighting the masked regions is shown in Figure~\ref{fig:f277w}.

We also produce a transmission spectrum using an independent reduction pipeline, \texttt{NAMELESS} \citep{feinstein2022, coulombe2022, radica2023}, as shown in Figure \ref{fig:ts_comparison}, to verify the self-consistency of our results (see Appendix \ref{sec:ind_pipeline}). The best-fitting transit parameters from NIRISS/SOSS white light curve fits are shown in Table~\ref{tab: WLC Parameters}.

\subsection{Occulted Starspot Method}
\begin{figure}
	\centering
	\includegraphics[width=\columnwidth]{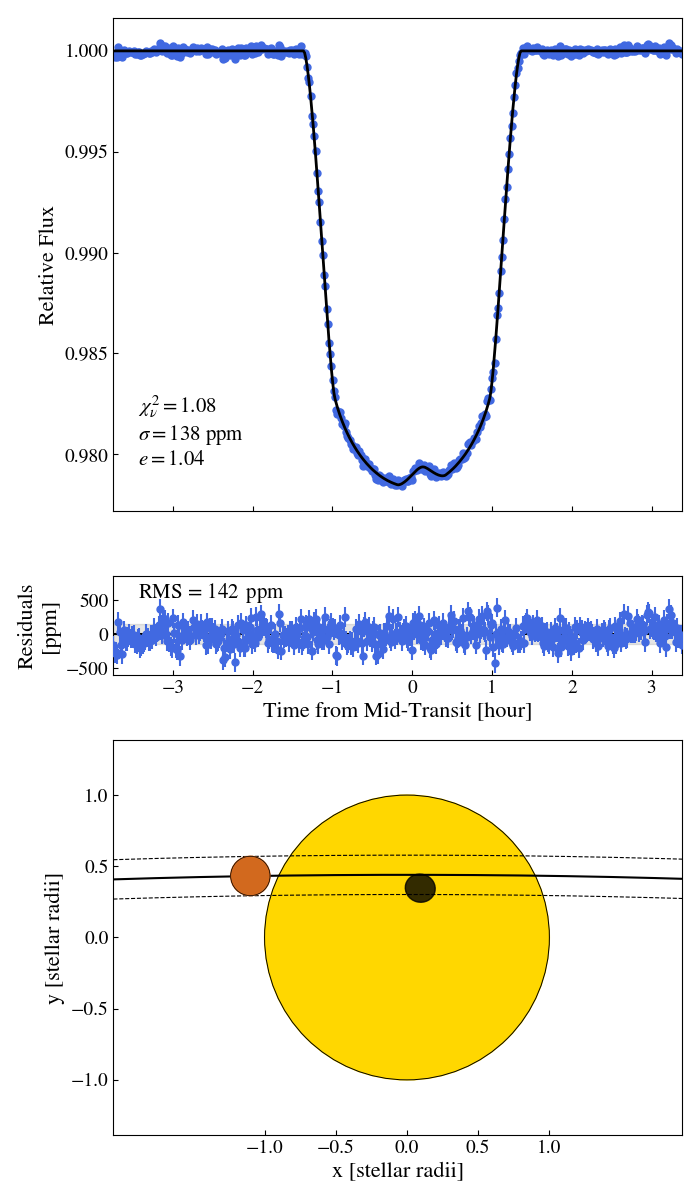}
    \caption{Modelling of the spot-crossing event. \emph{Top}: NIRISS/SOSS broadband light curve (blue), along with the best-fitting model overplotted (black) and the fit statistics listed in the bottom left corner ($\chi^2_{\nu}$; the reduced Chi-squared, $\sigma$; the average error bar, and \emph{e}; the error multiple needed to obtain a $\chi^2_{\nu}$ equal to unity). \emph{Middle}: Residuals to the transit fit with the RMS scatter. \emph{Bottom}: Physical representation of the maximum likelihood solution for the occulted starspot (black circle) on the star (yellow circle), along with the transit motion in black (dashed lines representing the transit chord) of the planet (orange circle). 
    \label{fig:starspot_lc}}
\end{figure}

We also perform a second light curve fit, enabling a joint inference of the starspot and planet properties to model the spot and study its impact on the transmission spectrum. To this end, we first compute a single broadband light curve by summing the flux from wavelengths 0.65--0.85\,$\mu$m of order 2 together with wavelengths 0.85--1.5\,$\mu$m of order 1. We exclude wavelengths bluewards of 0.65\,$\mu$m to maximize the S/N and wavelengths redwards of 1.5\,$\mu$m because the effect of spot crossings is stronger at shorter wavelengths where the spot contrast with respect to the stellar photosphere is larger. We fit a transit model with a spot-crossing event using \texttt{spotrod} \citep{beky2014}, which we have implemented into the \texttt{juliet} package \citep{espinoza2019b}. We fix the period, the eccentricity and the argument of periastron to the \citet{hartman2011} values and fit the mid-transit time, the impact parameter, the scaled semi-major axis, the scaled planet radius, the two parameters of the quadratic limb darkening law, and the jitter term with wide, uninformative priors. We also fit the zero point of the transit baseline, though the best-fitting value is again very close to zero. We fit four additional parameters to model the starspot: the $x$- and $y$-positions of the spot, its radius, and its spot-to-stellar flux contrast. The positions and the radius of the spot are in stellar radii units. The centre of the star is at (0, 0). We use uniform priors ranging from 0 to 1. We employ \texttt{dynesty} \citep{speagle2020} to sample the parameter space with 5000 live points. The reduced Chi-squared statistic for the fit with the highest likelihood is $\chi_\nu$ = 1.08. There is strong evidence ($\Delta\log Z$ = 95.7, or a $>$\,5\,$\sigma$ preference) for a transit model with an occulted spot. The best-fitting parameters for the broadband light curve fit with the spot-crossing modelled are displayed in Table \ref{tab: WLC Parameters}. The choice of wavelength range (0.65-1.5 $\mu$m) results in best-fitting values that are more precise and overall consistent at 1\,$\sigma$ to the best-fitting values with the entire wavelength ranges of order 1 and 2. We note that the $y$-position of the spot is not well constrained, as shown in the corner plot in Figure \ref{fig:corner_spot}, so we use the set of parameters with the highest likelihood instead of the medians of the posteriors from the broadband light curve fit, to ensure that the parameters are self-consistent. This best-fitting transit model is overplotted in the top panel of Figure \ref{fig:starspot_lc}, and a cartoon of the spot-crossing event is shown in the bottom panel. 

\begin{figure}
	\centering
	\includegraphics[width=0.85\columnwidth]{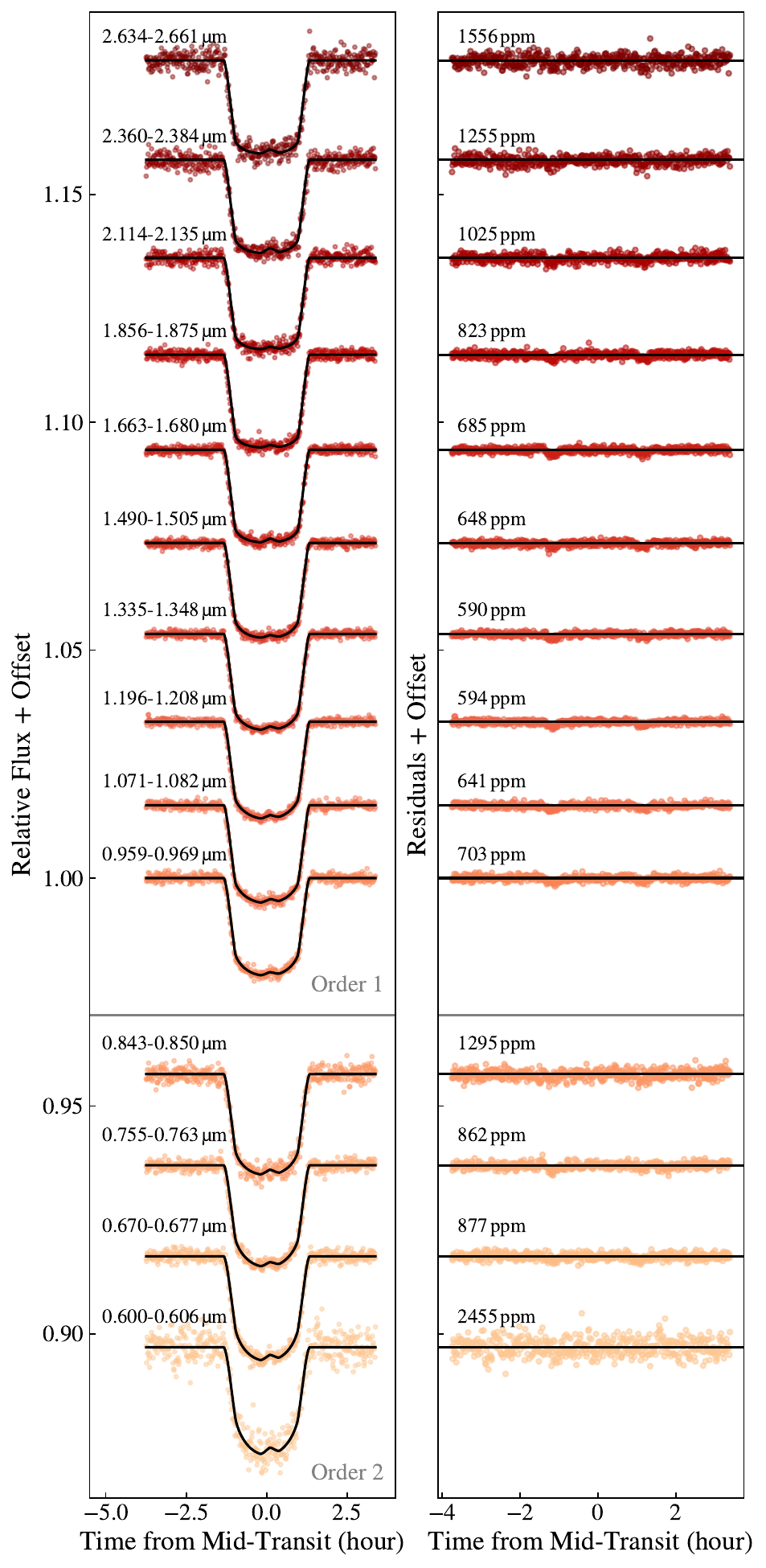}
    \caption{JWST NIRISS/SOSS binned spectrophotometric light curves, along with the best-fitting transit models using \texttt{spotrod} (black). \emph{Left}: Normalized spectrophotometric light curves at a resolving power of $R$ = 100. \emph{Right}: Associated residuals from the light curve fit in each bin with the RMS scatter indicated. 
    \label{fig:slc}}
\end{figure}

We then fit the spectrophotometric light curves at a resolving power of $R$ = 100, fixing the orbital parameters ($T_\textrm{0}$, $b$, $a/R_\textrm{*}$), as well as the spot positions and radius from the above broadband light curve fit. We fit the scaled planet radius, the limb-darkening parameters, the spot contrast and the jitter coefficient with the priors described above. The spectrophotometric light curves for 14 bins with their best-fitting transit models are shown in Figure \ref{fig:slc} for the highest likelihood model. The spot contrast is correlated with the $y$-position (see Figure \ref{fig:corner_spot}) because there is a degeneracy between the radius and the contrast of the spot given the strong correlation between the spot temperature and filling factor\footnote{The filling factor corresponds to the fraction of the planetary disc occulting the active region.} \citep{bruno2022}. To capture the uncertainties coming from the degeneracies in the spot $y$-position, size and contrast, we repeat the fit of the spectrophotometric light curves 50 times by fixing the same parameters ($T_\textrm{0}$, $b$, $a/R_\textrm{*}$, spot positions and radius) to 50 random sets taken from the posterior samples of the broadband light curve fit. We draw the 50 sets from the 68 \% of the weighted samples with the highest likelihoods.

We next constrain the temperature of the occulted spot by fitting PHOENIX stellar atmosphere model spectra \citep{husser2013} to the spot contrast spectrum resulting from the spectrophotometric light curves fits. Starspots' spectra are thought to be represented by stellar models with lower temperatures and 0.5--1\,dex lower $\log g$ than the host star. The increased magnetic pressure in active features decreases the gas pressure \citep{solanki2003,bruno2022}, which is akin to a stellar model with a lower surface gravity. In fitting for the spot temperature, we thus also left the spot surface gravity to vary as a free parameter. For the spot, we use stellar model spectra with temperatures from 4000 to 5000\,K, logarithmic surface gravities from 1 to 5\,dex and a fixed metallicity of 0.1 from \citet{hartman2011}. We interpolated these spectra in temperature and $\log{g}$ as needed during the fitting process. For the star, we use a stellar model spectrum with the temperature, $\log{g}$ and metallicity fixed to the corresponding values from \citet{hartman2011}. The model spot contrast spectrum is then simply the ratio of the model spot spectrum to the model star spectrum. We fit the temperature of the spot and the $\log{g}$ with wide, uninformative priors using the \texttt{emcee} MCMC package \citep{foreman-mackey2013}.

\subsection{Inferred Occulted Starspot Properties on HAT-P-18}
\begin{figure}
	\centering
	\includegraphics[width=\columnwidth]{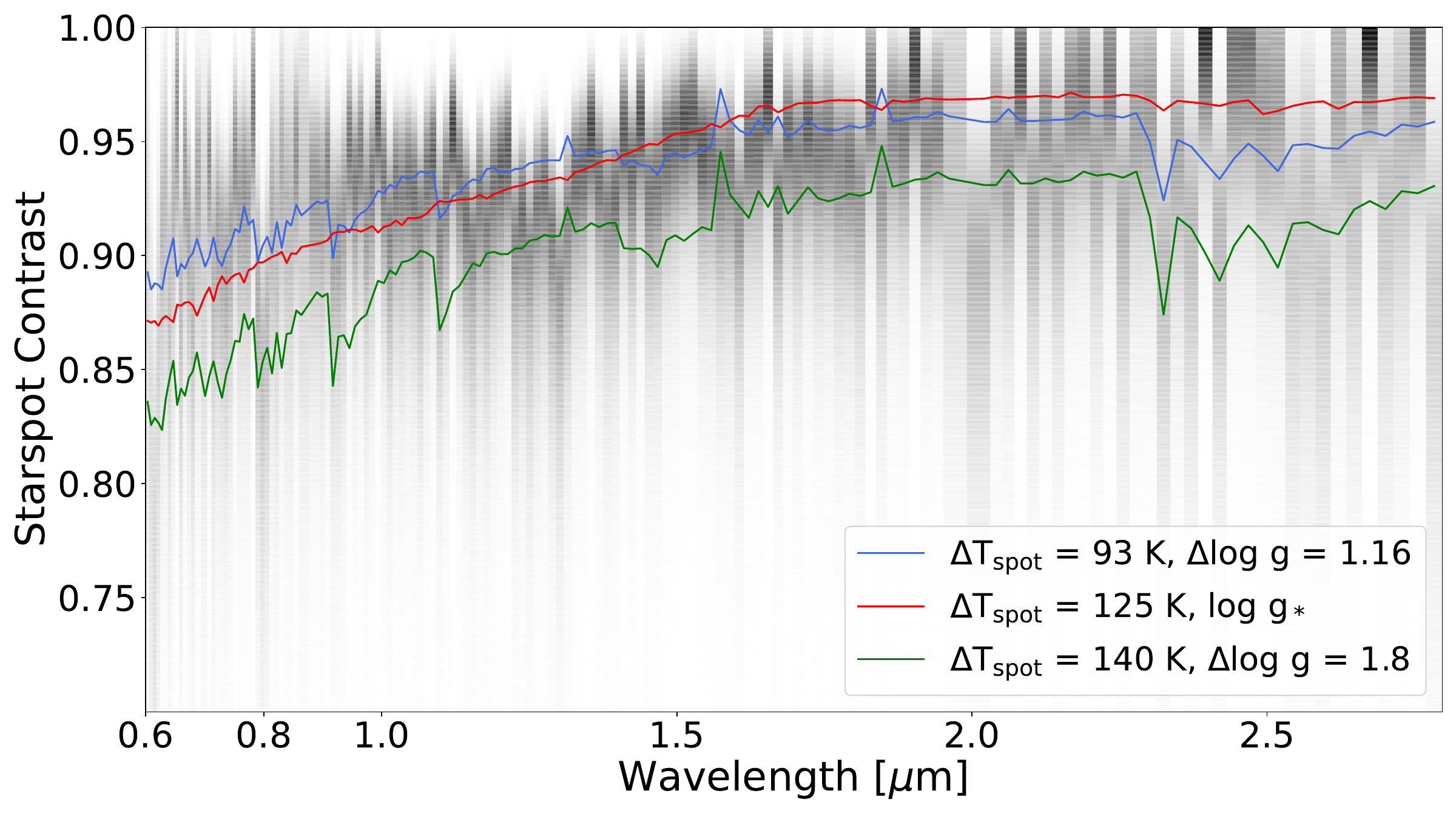}
    \caption{Density of contrast spectra from the 50 models fit, along with the most likely model in blue and a colder solution model in green (the one with a similar spot position). The model in red represents the best-fitting model to the highest likelihood contrast spectrum if we only fit the 
    spot temperature, fixing the surface gravity to the stellar value. The Chi-squared statistic for this fit is $\chi^2$ = 193 instead of $\chi^2$ = 156 when the spot model's surface gravity is also fit. 
    \label{fig:density}}
\end{figure}
From the aforementioned process, we obtained 51 starspot contrast spectra from the 51 spectrophotometric fits performed: one from the highest likelihood values and 50 with random draws from the broadband light curve fit. Each model has its own set of parameters (i.e., $T_\textrm{0}$, $b$, $a/R_\textrm{*}$, spot positions and radius). The resulting density of spot contrast spectra is shown in Fig. \ref{fig:density}. By inspecting the individual contrast spectra, we see that 28 of the 50 random fitting models (56\,\%) have a retrieved contrast spectrum within 1\,$\sigma$ of the one retrieved with the highest likelihood set of parameters. We lump these solutions together as the ``most likely'' one (blue model in Figure~\ref{fig:density}). The mean and standard deviation of the spot $x$- and $y$-position, radius, temperature, and the model $\log{g}$ are given in Table \ref{tab: familyspot} for this group of solutions. This corresponds to a spot radius of 0.116 $\pm$ 0.014 stellar radii, a temperature $\Delta T$ = -93 $\pm$ 15 K colder than the star (or \textit{T}\textsubscript{spot} = 4710 $\pm$ 15 K) and a lower $\log{g}$ by 1.16 $\pm$ 0.19\,dex. The retrieved $\log{g}$ for the starspot model is consistent with \citet{bruno2022} expectation.

Among the remaining contrast spectra, those not within 1\,$\sigma$ of the one with the highest likelihood, we can identify three other broad families of solutions corresponding to a colder spot because of a smaller filling factor; see Table \ref{tab: familyspot}. The transmission spectra for these four families of solutions do not significantly change (they are consistent at the 1\,$\sigma$ level) from the transmission spectrum obtained with the spot occultation features masked (as done in Section \ref{sec:lightcurve_mask}). This is expected for a spot of that size and temperature, as the TLSE from occulting such a spot results only in a small slope toward bluer wavelengths in the transmission spectrum, as shown in Figure \ref{fig:occulted_spot}. This effect, estimated to be of the order of 10 to 20 ppm from Equation \ref{eq:stellar_contam_factor_one_het} in Section \ref{sec:retrievals}, is smaller than our transit depth uncertainties. 
\begin{table*}
\caption{Families of solutions for the occulted starspot}
\label{tab: familyspot}
\begin{threeparttable}
    \begin{tabular}{ccccccc}
        \toprule
        \multirow{2}{*}{Solutions} & Fraction & x-position & y-position & size  & $\Delta$T & $\Delta$log g \\
        & of models &[$\rm R_*$] & [$\rm R_*$] & [$\rm R_*$] & [K] & [log$_{10}$ cm/s$^2$]\\
        \midrule 	
        \textbf{Most likely} & 0.56 &0.090 $\pm$ 0.005  & 0.42 $\pm$ 0.05 &  0.116 $\pm$ 0.014 & 93 $\pm$ 15 & 1.16 $\pm$ 0.19\\ \\ 
        \textbf{Colder spot} \\
        Similar position &0.18 & 0.092 $\pm$ 0.003  & 0.41 $\pm$ 0.07 & 0.090 $\pm$ 0.004 & 140 $\pm$ 20 & 1.8 $\pm$ 0.2 \\ \\ 
        Higher position &0.14& 0.088 $\pm$ 0.004  & 0.53 $\pm$ 0.03 & 0.125 $\pm$ 0.018 & 180 $\pm$ 50 & 2.1 $\pm$ 0.4 \\ \\
        Lower position &0.12& 0.091 $\pm$ 0.005  & 0.24 $\pm$ 0.05 & 0.14 $\pm$ 0.03 & 190 $\pm$ 80 & 2.2 $\pm$ 0.6\\ 
        \bottomrule
    \end{tabular}
\end{threeparttable}
\end{table*}

\section{Retrieval Analysis} \label{sec:retrievals}

We now present the inferences from our retrieval analysis of HAT-P-18\,b's transmission spectrum with NIRISS/SOSS (from Section \ref{sec:lightcurve_mask}). Our retrievals jointly model the influence of the planetary atmosphere and unocculted stellar heterogeneities on the transmission spectrum and thus yield simultaneous constraints. We conducted this analysis with three independent retrieval codes to ensure robust results.

In what follows, we first describe our approach for joint retrievals of a planetary atmosphere and stellar contamination. We then describe the setup of our three retrieval codes before presenting our resulting constraints on HAT-P-18\,b's atmosphere and unocculted stellar active regions. Finally, we compare retrieval results from our HST + \textit{Spitzer} transmission spectrum to those yielded from NIRISS/SOSS.

\subsection{A Joint Planetary Atmosphere and Stellar Contamination Retrieval Method}

The influence of active stellar regions on exoplanet transmission spectra --- the \emph{transit light source effect} \citep[e.g.,][]{rackham2018} --- can be modelled simultaneously with the planetary atmosphere in retrievals. This has the advantage of yielding joint constraints on the planetary atmosphere and unocculted stellar heterogeneities while also propagating uncertainties from the star into the derived atmospheric constraints \citep[see][for a review]{rackham2023a}. Several retrieval studies have included prescriptions for unocculted starspots and/or faculae \citep[e.g.,][]{pinhas2018,bixel2019,iyer2020,rathcke2021,jiang2021,rackham2023b,thompson2023}, with the most common being a three-parameter model that fits for the temperature and covering fraction of a single heterogeneity \citep{pinhas2018}. While these treatments have generally sufficed for \emph{Hubble} and ground-based data, the exceptional data quality and wavelength coverage of JWST motivates the consideration of more complex stellar contamination models.

We investigate a range of unocculted stellar heterogeneity prescriptions during our HAT-P-18\,b retrievals. Our goal is to determine the appropriate level of starspot and/or faculae model complexity required to interpret the JWST/NIRISS transmission spectrum of HAT-P-18\,b while also informing future JWST retrieval studies of hot Jupiters. We begin by summarizing the equations underlying the transit light source effect. 

\subsubsection{The Transit Light Source Effect}\label{sec:TLS_descr}

The transmission spectrum of an exoplanet transiting a star with a heterogeneous stellar surface can be written as \citep{macdonald2022}
\begin{equation}
    \Delta_{\lambda} = \delta_{\lambda, \, \rm{atm}} \, \epsilon_{\lambda, \, \rm{het}} \, \psi_{\lambda, \, \rm{night}}
\label{eq:transmission_spectrum_joint}
\end{equation}
where $\Delta_{\lambda}$ is the observed transmission spectrum, $\delta_{\lambda, \, \rm{atm}}$ is the transmission spectrum from the planetary atmosphere, $\epsilon_{\lambda, \, \rm{het}}$ is the wavelength-dependent ``stellar contamination factor'' from a heterogeneous stellar surface, and $\psi_{\lambda, \, \rm{night}}$ accounts for thermal emission from the planetary nightside (we assume this final term negligibly deviates from unity for HAT-P-18\,b). For a planetary atmosphere with properties varying only with altitude (the 1D assumption), $\delta_{\lambda, \, \rm{atm}}$ is given by        
\begin{equation}
    \delta_{\lambda, \, \rm{atm}} = \frac{R_{\mathrm{p, \, top}}^2 - 2 \displaystyle\int_{0}^{R_{\mathrm{p, \, top}}} b \, e^{-\tau_{\lambda, \, \rm{slant}} (b)} \, db}{R_{\mathrm{*}}^{2}}
\label{eq:atm_transmission_spectrum}  
\end{equation} 
where $R_{\mathrm{p, \, top}}$ is the planetary radius at the top of the modelled atmosphere, $b$ is the ray impact parameter, $R_*$ is the stellar radius, and $\tau_{\lambda, \, \rm{slant}}$ is the slant optical depth. For a stellar surface with $N_{\rm{het}}$ unocculted heterogeneous active regions (e.g., spots or faculae), the stellar contamination factor, $\epsilon_{\lambda, \, \rm{het}}$, can be expressed as
\begin{equation}
    \epsilon_{\lambda, \, \rm{het}} = \frac{1}{1 - \displaystyle\sum_{i=1}^{N_{\rm{het}}} f_{\mathrm{het}, \, i} \left(1 - \frac{I_{\lambda, \, \mathrm{het}, \, i}}{I_{\lambda, \, \rm{phot}}} \right)}
\label{eq:stellar_contam_factor}   
\end{equation}
where $f_{\mathrm{het}, \, i}$ is the fractional stellar disc coverage of the $i^{\rm{th}}$ heterogeneous region, with corresponding specific intensity $I_{\lambda, \, \mathrm{het}, \, i}$, while $I_{\lambda, \, \rm{phot}}$ is the specific intensity of the stellar photosphere. Equation~\ref{eq:stellar_contam_factor} demonstrates that the stellar contamination factor deviates from unity when a heterogeneity possesses a different intensity from the photosphere. In general, $I_{\lambda, \, \mathrm{het}, \, i}$ is shorthand for $I_{\lambda, \, \mathrm{het}, \, i} \, (\mathrm{[Fe/H]}_i, \mathrm{log} \, g_i, T_i)$, where $\mathrm{[Fe/H]}_i$ is the local metallicity of the heterogeneity, $\mathrm{log} \, g_i$ is the local log surface gravity, and $T_i$ is the heterogeneity temperature.

Most retrieval studies consider a single population of unocculted heterogeneities with a temperature different from that of the photosphere (but with the same metallicity and surface gravity). Given these assumptions, the stellar contamination factor is given by \citep[e.g.,][]{rackham2018,pinhas2018} 
\begin{equation}
    \epsilon_{\lambda, \, \rm{het}} = \frac{1}{\displaystyle 1 -  f_{\mathrm{het}} \left(1 - \frac{I_{\lambda, \, \mathrm{het}} \, (T_{\rm{het}})}{I_{\lambda, \, \rm{phot}} \, (T_{\rm{phot}})} \right)}
\label{eq:stellar_contam_factor_one_het}   
\end{equation}
where the three free parameters are the heterogeneity coverage fraction, $f_{\mathrm{het}}$, the heterogeneity temperature, $T_{\rm{het}}$, and the photosphere temperature, $T_{\rm{phot}}$.

We also consider here a more general treatment of stellar contamination. Motivated by several strategic gaps identified in the NASA ExoPAG SAG 21 community report \citep{rackham2023a}, we also consider a two-heterogeneity model with the assumption of a common surface gravity relaxed. By defining these heterogeneities as starspots and faculae ($T_{\rm{spot}}$ < $T_{\rm{phot}}$ < $T_{\rm{fac}}$), we can write the stellar contamination factor as
\begin{equation}
\begin{split}
 \epsilon_{\lambda, \, \mathrm{het}} = \frac{1}{ \displaystyle 1 - \biggl[ f_{\mathrm{spot}} \left(1 - \frac{I_{\lambda, \, \mathrm{spot}} \, (\mathrm{log} \, g_{\mathrm{spot}}, \, T_{\mathrm{spot}})}{I_{\lambda, \, \mathrm{phot}} \, (\mathrm{log} \, g_{\mathrm{phot}}, \, T_{\mathrm{phot}})} \right) + } \\
    { f_{\mathrm{fac}} \left(1 - \frac{I_{\lambda, \, \mathrm{fac}} \, (\mathrm{log} \, g_{\mathrm{fac}}, \, T_{\mathrm{fac}})}{I_{\lambda, \, \mathrm{phot}} \, (\mathrm{log} \, g_{\mathrm{phot}}, \, T_{\mathrm{phot}})} \right) \biggr] }
\end{split}
\label{eq:stellar_contam_factor_two_het}   
\end{equation}
The addition of a second heterogeneity population and separate surface gravities increases the number of free parameters for this stellar contamination model to eight. 

\subsubsection{Unocculted Stellar Heterogeneity Models for HAT-P-18 b} \label{sec:TLS_models}

We consider five treatments for unocculted stellar heterogeneities when retrieving HAT-P-18\,b's transmission spectrum:

\begin{enumerate}
    \item \textbf{No stellar contamination}.
    \item \textbf{One heterogeneity}, defined by three free parameters:  $f_{\mathrm{het}}$, $T_{\rm{het}}$, and $T_{\rm{phot}}$.
    \item \textbf{Two heterogeneities} (spots + faculae), defined by five free parameters: $f_{\mathrm{spot}}$, $T_{\rm{spot}}$, $f_{\mathrm{fac}}$, $T_{\rm{fac}}$, and $T_{\rm{phot}}$.
    \item \textbf{One heterogeneity with free surface gravity}, defined by five free parameters: $f_{\mathrm{het}}$, $T_{\rm{het}}$, $\mathrm{log} \, g_{\mathrm{het}}$, $T_{\rm{phot}}$, and $\mathrm{log} \, g_{\mathrm{phot}}$. 
    \item \textbf{Two heterogeneities with free surface gravity}, defined by eight free parameters: $f_{\mathrm{spot}}$, $T_{\rm{spot}}$, $\mathrm{log} \, g_{\mathrm{spot}}$, $f_{\mathrm{fac}}$, $T_{\rm{fac}}$, $\mathrm{log} \, g_{\mathrm{fac}}$, $T_{\rm{phot}}$, and $\mathrm{log} \, g_{\mathrm{phot}}$.
\end{enumerate}

\subsection{Retrieval Configuration} \label{sec:retrieval_config}


We applied three exoplanet retrieval codes to HAT-P-18\,b's transmission spectrum: \textsc{Poseidon} \citep{macdonald2017,macdonald2023}, \textsc{Aurora} \citep{welbanks2021}, and \texttt{SCARLET} \citep{benneke2012,benneke2015}. We initially conducted independent analyses to establish which atmospheric and stellar properties provide the necessary complexity to describe HAT-P-18\,b's transmission spectrum. After comparing our results, we chose a common set of model assumptions for the three codes: chemical opacity from Na, K, H$_2$O, CO, CO$_2$, CH$_4$, HCN, and NH$_3$ (the most prominent opacity sources expected at HAT-P-18\,b's equilibrium temperature under equilibrium chemistry and/or with chemical quenching, see e.g., \citealt{madhusudhan2016}); an inhomogeneous grey cloud-deck combined with a power-law haze; and one unocculted stellar heterogeneity (i.e., the \textit{one heterogeneity} model). Our retrievals were conducted on the $R$ = 100 binned variant of HAT-P-18\,b's transmission spectrum with the spot-crossing masked (see Figure~\ref{fig:transmission_spectra}), but we found consistent results when retrieving the full-resolution NIRISS data. We describe the specific configuration used by each retrieval code below and list the priors used in Table~\ref{tab:retrieval_priors}.

\newcommand{\ra}[1]{\renewcommand{\arraystretch}{#1}}
\begin{table}
    \ra{1.2}
    \caption{Retrieval parameters and priors.}
    \begin{tabular*}{\columnwidth}{l@{\extracolsep{\fill}} lllll@{}}\toprule
    Parameter & \textsc{Poseidon} & \texttt{SCARLET} & \textsc{Aurora} \\ \midrule
    \textbf{Composition} & \\ 
    \hspace{0.5em} $\log X_{i}$ & $\mathcal{U}$\,[-12, -1] & $\mathcal{U}$\,[-12, -0.3] & $\mathcal{U}$\,[-12, -1] \\ \midrule
    \textbf{P-T profile} & \\ 
    \hspace{0.5em} $\alpha_{1,2}$ & $\mathcal{U}$\,[0.02, 2.0] & --- & $\mathcal{U}$\,[0.02, 2.0]  \\
    \hspace{0.5em} $\log P_{1,2}$ & $\mathcal{U}$\,[-8, 2] & --- & $\mathcal{U}$\,[-8, 2] \\
    \hspace{0.5em} $\log P_{3}$ & $\mathcal{U}$\,[-2, 2] & --- & $\mathcal{U}$\,[-2, 2] \\
    \hspace{0.5em} $T_{\mathrm{ref}}$ & $\mathcal{U}$\,[300, 2000] & $\mathcal{U}$\,[300, 1700] & $\mathcal{U}$\,[300, 1000] \\ \midrule
    \textbf{Aerosols} & \\
    \hspace{0.5em} $\log c_\mathrm{haze}$ & --- & $\mathcal{U}$\,[-10, 5] & --- \\
    \hspace{0.5em} $\log a$ & $\mathcal{U}$\,[-4, 8] & --- & $\mathcal{U}$\,[-4, 8] \\
    \hspace{0.5em} $\gamma$ & $\mathcal{U}$\,[-20, 2] & --- & $\mathcal{U}$\,[-20, 2] \\
    \hspace{0.5em} $\log P_{\mathrm{cloud}}$ & $\mathcal{U}$\,[-8, 2] & $\mathcal{U}$\,[-8, 2] & $\mathcal{U}$\,[-8, 2] \\
    \hspace{0.5em} $f_{\mathrm{cloud}}$ & $\mathcal{U}$\,[0, 1] & $\mathcal{U}$\,[0, 1] & $\mathcal{U}$\,[0, 1] \\ \midrule
    \textbf{Stellar} & \\
    \hspace{0.5em} \textbf{One het.} \\
    \hspace{1.0em} $f_{\mathrm{het}}$ & $\mathcal{U}$\,[0.0, 0.5] & $\mathcal{U}$\,[0.0, 1.0] & $\mathcal{U}$\,[0.0, 0.5] \\
    \hspace{1.0em} $T_{\mathrm{het}}$ & $\mathcal{U}$\,[3500, 1.2\,$T_{*}$] & --- & $\mathcal{U}$\,[3500, 1.2\,$T_*$] \\
    \hspace{1.0em} $\Delta T_{\mathrm{het}}$ & --- & $\mathcal{U}$\,[-800, -50] & --- \\
    \hspace{1.0em} $T_{\mathrm{phot}}$ & $\mathcal{N}$\,[$T_{*}$, $\sigma_{T_{*}}$] & $\mathcal{N}$\,[$T_{*}$, $\sigma_{T_{*}}$] & $\mathcal{N}$\,[$T_{*}$, $\sigma_{T_{*}}$] \\
    \hspace{1.0em} $\mathrm{log} \, g_{\mathrm{het}}$ & $\mathcal{U}$\,[3.0, 5.0] & --- & --- \\
    \hspace{1.0em} $\mathrm{log} \, g_{\mathrm{phot}}$ & $\mathcal{N}$ [$\log g_{*}$, $\sigma_{\log g_{*}}$] & --- & --- \\
    \hspace{0.5em} \textbf{Two het.} \\
    \hspace{1.0em} $f_{\mathrm{fac}}$ & $\mathcal{U}$\,[0.0, 0.5] & --- & --- \\
    \hspace{1.0em} $f_{\mathrm{spot}}$ & $\mathcal{U}$\,[0.0, 0.5] & --- & --- \\
    \hspace{1.0em} $T_{\mathrm{fac}}$ & $\mathcal{U}$ [$T_{*} - 3\,\sigma_{T_{*}}$, 1.2\,$T_{*}$] & --- & --- \\
    \hspace{1.0em} $T_{\mathrm{spot}}$ & $\mathcal{U}$ [3500, $T_{*} + 3\,\sigma_{T_{*}}$] & --- & --- \\
    \hspace{1.0em} $T_{\mathrm{phot}}$ & $\mathcal{N}$\,[$T_{*}$, $\sigma_{T_{*}}$] & --- & --- \\
    \hspace{1.0em} $\mathrm{log} \, g_{\mathrm{fac}}$ & $\mathcal{U}$\,[3.0, 5.0] & --- & --- \\
    \hspace{1.0em} $\mathrm{log} \, g_{\mathrm{spot}}$ & $\mathcal{U}$\,[3.0, 5.0] & --- & --- \\
    \hspace{1.0em} $\mathrm{log} \, g_{\mathrm{phot}}$ & $\mathcal{N}$ [$\log g_{*}$, $\sigma_{\log g_{*}}$] & --- & --- \\ \midrule
    \textbf{Other} & \\
    \hspace{0.5em} $R_{\mathrm{p, \, ref}}$ & $\mathcal{U}$\,[0.85\,$R_{\mathrm{p}}$, 1.15\,$R_{\mathrm{p}}$] & --- & --- \\
    \hspace{0.5em} $\log P_{\mathrm{ref}}$ & --- & --- & $\mathcal{U}$\,[-8, 2] \\
    \bottomrule
    \vspace{-0.5pt}
    \end{tabular*}
    \textit{Note:} All three retrieval codes adopt the \textit{one heterogeneity} stellar contamination model for this comparison. $T_{\mathrm{ref}}$ is defined at 10\,mbar for \textsc{Poseidon}, 10$^{-8}$\,bar for \textsc{Aurora}, and is the isothermal temperature for \texttt{SCARLET}. All pressure parameters are expressed in units of bar and temperatures in K. For the priors, we adopt $R_{\rm{p}} = 0.995\,R_J$, $T_{*} = 4803$\,K, $\log g_{*} = 4.57$ (cgs), $\sigma_{T_{*}} = 80$\,K, and $\sigma_{\log g_{*}} = 0.04$ (cgs). All retrievals have a fixed stellar metallicity of [Fe/H] = 0.1, while those without free surface gravity assume $\log g = \log g_{*} = 4.57$ in all regions. `---' indicates that the parameter did not feature in the retrieval. All log parameters are base 10. 
    \label{tab:retrieval_priors}
\end{table}

\subsubsection{\textsc{Poseidon}}

We conducted a series of \textsc{Poseidon} retrievals to assess the model complexity required to fit HAT-P-18\,b's JWST/NIRISS transmission spectrum. This includes the five stellar contamination models listed in Section~\ref{sec:TLS_models}, nested models to compute detection significances for each chemical species, and additional robustness tests (e.g., a retrieval of the unbinned pixel-resolution spectrum). 

Our \textsc{Poseidon} retrievals model HAT-P-18\,b has a H$_2$+He-dominated atmosphere with a standard configuration used for hot Jupiter retrieval studies \citep[e.g.,][]{macdonald2017,kirk2021,taylor2023}. We prescribe 100 layers, spaced uniformly in log-pressure from $10^{-8}$--$100$\,bar, where the layers follow a variant of the pressure-temperature (P-T) profile parameterization from \citet{madhusudhan2009} but with the reference temperature located at 10\,mbar. The inhomogeneous cloud and haze aerosol model follows \citet{macdonald2017}. We fit for the $\log_{10}$ volume mixing ratios of the 8 gases in the reference model, using absorption cross-sections \citep[see][]{macdonald2022} computed from the following line lists: Na and K \citep{ryabchikova2015}; H$_2$O \citep{polyansky2018}; CO \citep{li2015}; CO$_2$ \citep{tashkun2011}; CH$_4$ \citep{yurchenko2017}; HCN \citep{barber2014}; and NH$_3$ \citep{Coles2019}. We include H$_2$-H$_2$ collision-induced absorption from HITRAN \citep{karman2019}. We also fit for the planetary radius at a reference pressure of 10\,bar. We calculate model transmission spectra from 0.55--2.9\,$\mu$m at $R$ = 20,000. We calculate the stellar contamination factor (Equation~\ref{eq:stellar_contam_factor}) by interpolating PHOENIX models \citep{husser2013} using the \texttt{PyMSG} package \citep{townsend2023}. Our \textsc{Poseidon} retrievals have up to 27 free parameters (depending on the chosen stellar contamination model from Section~\ref{sec:TLS_models}), with 1,000 nested sampling live points used by the \texttt{PyMultiNest} \citep{feroz2009,buchner2014} package to chart the parameter space.

\subsubsection{\texttt{SCARLET}}\label{sec:scarlet}

We performed retrievals using the \texttt{SCARLET} retrieval framework \citep{benneke2015, benneke2019a}. We improved the \texttt{SCARLET} framework from previous work on low-resolution observations \citep{benneke2019a, benneke2019b, piaulet2023} with the addition of non-uniform cloud coverage and of the contribution of stellar contamination to the transmission spectrum. 

In the \texttt{SCARLET} retrieval, the atmosphere is parameterized by the abundances of the spectrally-active chemical species of interest (Na, K, H$_2$O, CO, CO$_2$, CH$_4$, HCN, NH$_3$) which are assumed to be well-mixed and constant with pressure, and an isothermal temperature structure. The ratio of He and H$_2$, which make up the remainder of the gas, is set to Jupiter's He/H$_2$ of 0.157.

The retrieval includes three parameters describing aerosols ($\log p_\mathrm{cloud}$, $f_\mathrm{cloud}$, $\log c_\mathrm{haze}$). The cloud parameterization consists of a grey cloud, opaque across all wavelengths, with a cloud top pressure $p_\mathrm{cloud}$. Potential fractional cloud coverage is captured by the $f_\mathrm{cloud}$ parameter and corresponds to a weighted average of a cloud-free and a cloudy model. The contribution of haze to the transmission spectrum via a slope towards shorter wavelengths is parameterized using the $c_\mathrm{haze}$ parameter, which is a factor that multiplies the Rayleigh contribution to the scattering coefficient (unity corresponds to standard Rayleigh scattering).

Finally, two parameters describe the stellar heterogeneity: $\Delta T_\mathrm{spot}$ and $f_\mathrm{spot}$. Stellar heterogeneity is implemented following the standard approach (see Section \ref{sec:TLS_descr}) assuming that the contribution of stellar contamination to the spectrum can be represented by that of spots with a temperature of $T_\mathrm{spot} = T_\mathrm{phot} + \Delta T_\mathrm{spot}$ (where $T_\mathrm{phot}$ is the photosphere temperature) covering a fraction $f_\mathrm{spot}$ of the star. Stellar models are queried from the PHOENIX grid of stellar atmosphere models \citep{husser2013}.

The forward models calculated by the retrieval have a resolving power of 15,625. We use opacity sampling from cross-section tables at a resolving power of 250,000 to build the opacity tables used, and the opacities are taken from the ExoClimes Simulation Platform database. We choose HITEMP opacities for H$_2$O, CO$_2$, CH$_4$, NH$_3$ \citep{rothman2010}, and ExoMol opacities for all other species \citep{tennyson2016}. We use Nested Sampling to sample the parameter space and its Python implementation in the \texttt{nestle} module \citep{skilling2004,mukherjee2006,shaw2007}.

\subsubsection{\textsc{Aurora}}\label{sec:aurora}

We also perform atmospheric retrievals with \textsc{Aurora} \citep{welbanks2021} to interpret the transmission spectrum of HAT-P-18\,b. The application of \textsc{Aurora} to transmission spectra of exoplanets is described in \citet{welbanks2022} following the general setup presented \citet{welbanks2019a, welbanks2019b}. We follow the methods of \citet{pinhas2018} to consider the impact of stellar heterogeneities as implemented in previous works \citep[see e.g.,][]{ahrer2023b}. 

Aurora computes line-by-line radiative transfer in transmission geometry for a parallel-plane atmosphere assuming hydrostatic equilibrium. Our atmospheric model is computed using a 100 layer pressure grid uniformly spaced in log-pressure between $10^{-8}$ and $100$~bar, and a wavelength grid from 0.55\,$\mu$m to 2.9\,$\mu$m, covering this NIRISS observation, at a constant resolution of $R$ = 20,000. We parameterize the P-T profile of the atmosphere using the prescription from \citet{madhusudhan2009} and the presence of inhomogeneous clouds \citep[e.g.,][]{line2016} and hazes \citep[e.g.,][]{lecavelier2008}, using a single sector for their combined spectroscopic imprint following the description in \citet{welbanks2021}. The volume mixing ratios for the eight chemical species considered in our fiducial model are assumed to be constant with height and use cross-sections for H$_2$–-H$_2$ and H$_2$–-He collision induced absorption \citep[CIA;][]{richard2012}, CH$_4$ \citep{yurchenko2014}, CO \citep{rothman2010}, CO$_2$ \citep{rothman2010}, H$_2$O \citep{rothman2010}, HCN \citep{barber2014}, K \citep{allard2016}, Na \citep{allard2019}, and NH$_3$ \citep{yurchenko2011}, computed as described in \citet{gandhi2017,gandhi2018} and \citet{welbanks2019b}. Following \citet{pinhas2018}, we model the heterogeneous stellar photosphere by interpolating stellar models from the PHOENIX database \citep{husser2013}. The parameter estimation is performed using the nested sampling algorithm \citep{skilling2004} with MultiNest \citep{feroz2009} and its python implementation PyMultiNest \citep{buchner2014}. 

\subsection{An Explanation for HAT-P-18 b's Transmission Spectrum} \label{sec:results_spectrum_model}

\begin{figure}
    \centering
    \includegraphics[width=\columnwidth]{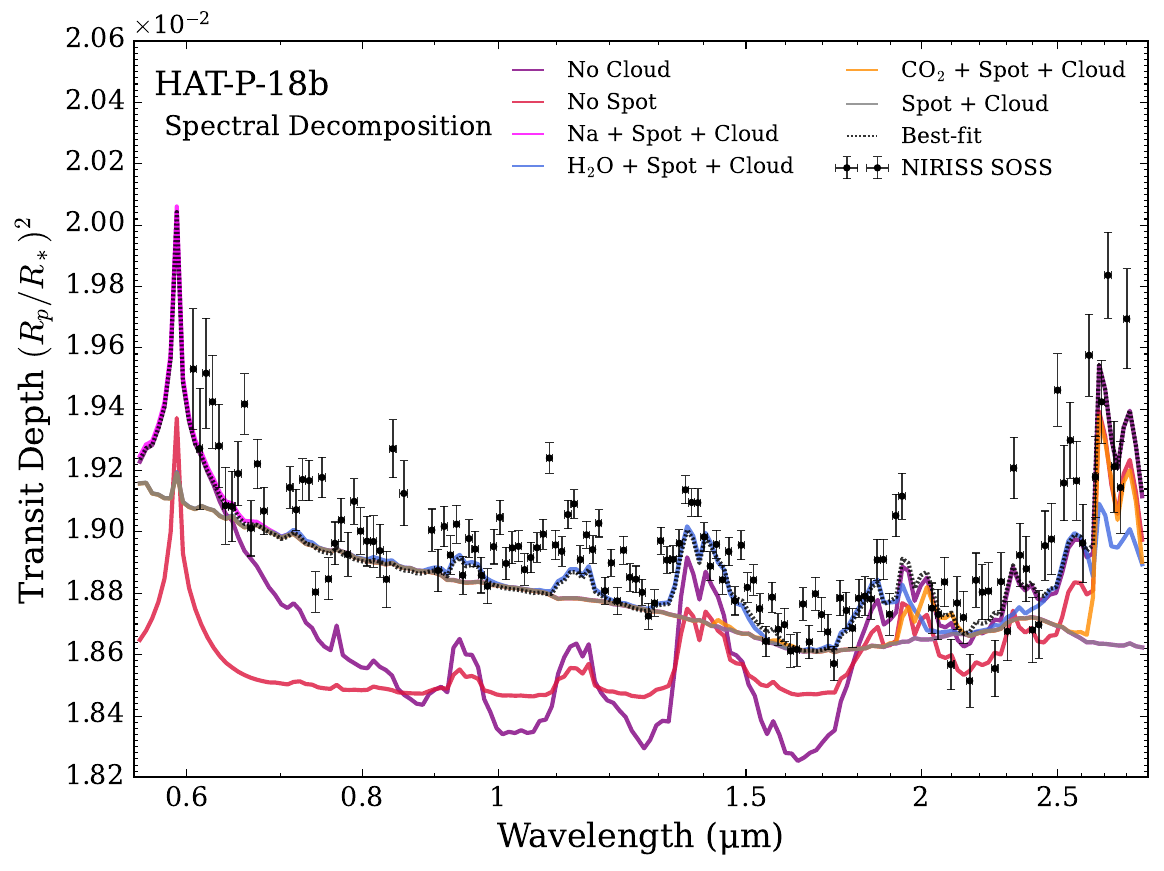}
    \caption{Contributions of planetary atmosphere and stellar features to HAT-P-18\,b's transmission spectrum. The best-fitting model spectrum from the \textsc{Poseidon} \textit{one heterogeneity} retrieval (dashed line) --- see Section~\ref{sec:results_atmo} --- is compared to the NIRISS/SOSS data (error bars) for reference. The coloured lines show the effect of removing the cloud (purple) and starspot (red) from the best-fitting model, while the other lines show models with combinations of a cloud, starspot, H$_2$ continuum opacity, and a single atmospheric chemical species (Na: magenta; H$_2$O: blue; CO$_2$: orange) or no additional absorbers (grey). HAT-P-18\,b's transmission spectrum is therefore well-explained by unocculted starspots, a cloud deck, Na, H$_2$O, and CO$_2$.}
\label{fig:spectral_decomposition}
\end{figure}

We first provide an intuitive explanation for HAT-P-18\,b's transmission spectrum. Figure~\ref{fig:spectral_decomposition} shows a spectral decomposition of the maximum likelihood (best-fitting) model transmission spectrum from the \textsc{Poseidon} \textit{one heterogeneity} retrieval (model (ii) in Section~\ref{sec:TLS_models}). We detect multiple H$_2$O absorption features near 0.95\,$\mu$m, 1.15\,$\mu$m, 1.4\,$\mu$m, 1.9\,$\mu$m, and 2.6\,$\mu$m (with a combined significance of 12.5\,$\sigma$). We further detect a CO$_2$ absorption feature near 2.7\,$\mu$m (7.3\,$\sigma$) and infer evidence of Na at lower significance (2.7\,$\sigma$). The small amplitude of the absorption features is due to an optically thick cloud deck which is uniform around the terminator (7.4\,$\sigma$). The inference of non-patchy clouds is driven by the wing shape of molecular bands \citep{macdonald2017} --- especially the H$_2$O band centred near 1.4\,$\mu$m --- and the flat continuum from 1.6--1.8\,$\mu$m. Finally, the spectral slope shortwards of 1.65\,$\mu$m is caused by the combination of unocculted starspots (5.8\,$\sigma$) and the uniform cloud deck. The quoted detection significances are from Bayesian model comparisons between the reference \textsc{Poseidon} \textit{one heterogeneity} model and another retrieval with one model component removed \citep[e.g.,][]{benneke2013}. We do not identify statistically significant evidence for K, CO, CH$_4$, HCN, NH$_3$, nor a scattering haze. We note that these conclusions are independently retrieved by the \texttt{SCARLET} and \textsc{Aurora} retrievals, which produce similar best-fitting spectra (see Section~\ref{sec:results_atmo}). Finally, we determine below that a single high data point near 1.1\,$\mu$m, not fit by our retrieval models, is attributable to metastable helium. We verified that removing this data point does not alter the retrieval results.

\subsubsection{Evidence of Helium Absorption}\label{sec:helium}

\begin{figure}
    \centering
    \includegraphics[width=0.96\columnwidth]{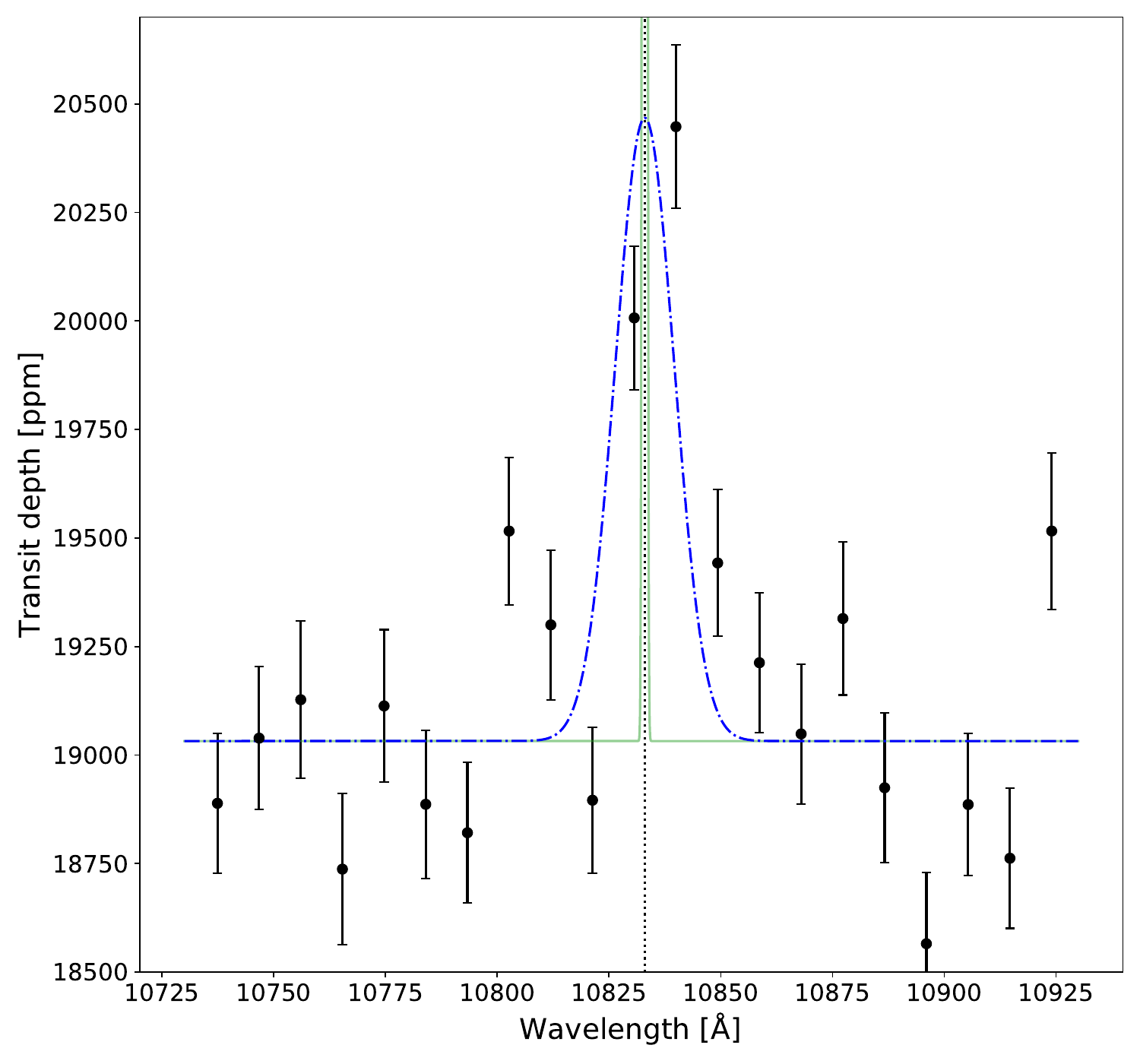}
    \caption{Pixel-resolution transmission spectrum of HAT-P-18\,b around the 1.083\,$\mu$m helium triplet. We model the helium triplet as a Gaussian convolved to the resolution of NIRISS/SOSS and fitted to the data. The best-fit is shown in blue, and the unconvolved Gaussian in green. The central wavelength of the metastable helium triplet is indicated as the vertical dashed line.
    \label{fig:helium}}
\end{figure}

We assessed evidence of helium absorption in HAT-P-18\,b's atmosphere via an independent fit to the pixel-resolution transmission spectrum near 1.083\,$\mu$m. Figure~\ref{fig:helium} shows that a Gaussian fit, with free amplitude and width, favours an excess absorption of 0.13\,$\pm$\,0.03\,\% (4.3\,$\sigma$) centred around the near-infrared helium triplet at 1.083\,$\mu$m. Our result is consistent within 1\,$\sigma$ to \citet{fu2022}, \citet{vissapragada2022}, and \citet{paragas2021}. Since the helium line profile is unresolved with NIRISS/SOSS, even at pixel resolution, we do not conduct more complex modelling of the helium line. Consequently, we cannot unambiguously attribute this signature to the planetary atmosphere nor disentangle the layers probed by the helium triplet. Nonetheless, we confirm that HAT-P-18\,b exhibits a longer transit duration with more absorption post transit --- as also observed by \citet{fu2022} --- which could indicate an extended atmosphere with the presence of a cometary-like tail as exhibited by WASP-107\,b \citep{allart2019}.

Future observations of HAT-P-18\,b at higher spectral resolution can confirm the helium absorption. A confirmation would indicate that HAT-P-18\,b is losing its atmosphere and is undergoing significant mass loss. We estimated the intrinsic excess absorption arising from the resolved helium triplet to be 3.17\,$\pm$\,0.67\,\% (Figure~\,\ref{fig:helium}). This value was derived by fitting a Gaussian with a fixed width of 1\,\AA\ (the typical width of the helium triplet; e.g., \citealt{allart2018, allart2019, nortmann2018, salz2018}) and a free intrinsic amplitude, which was then convolved to the pixel-resolution NIRISS/SOSS data. Such a strong helium signature would be readily detectable with ground-based high-resolution spectrographs, which would provide complementary observations to study the unique processes shaping HAT-P-18\,b's upper atmosphere.

\subsection{The Atmosphere of HAT-P-18 b} \label{sec:results_atmo}

\begin{figure*}
    \centering
    \includegraphics[width=\textwidth]{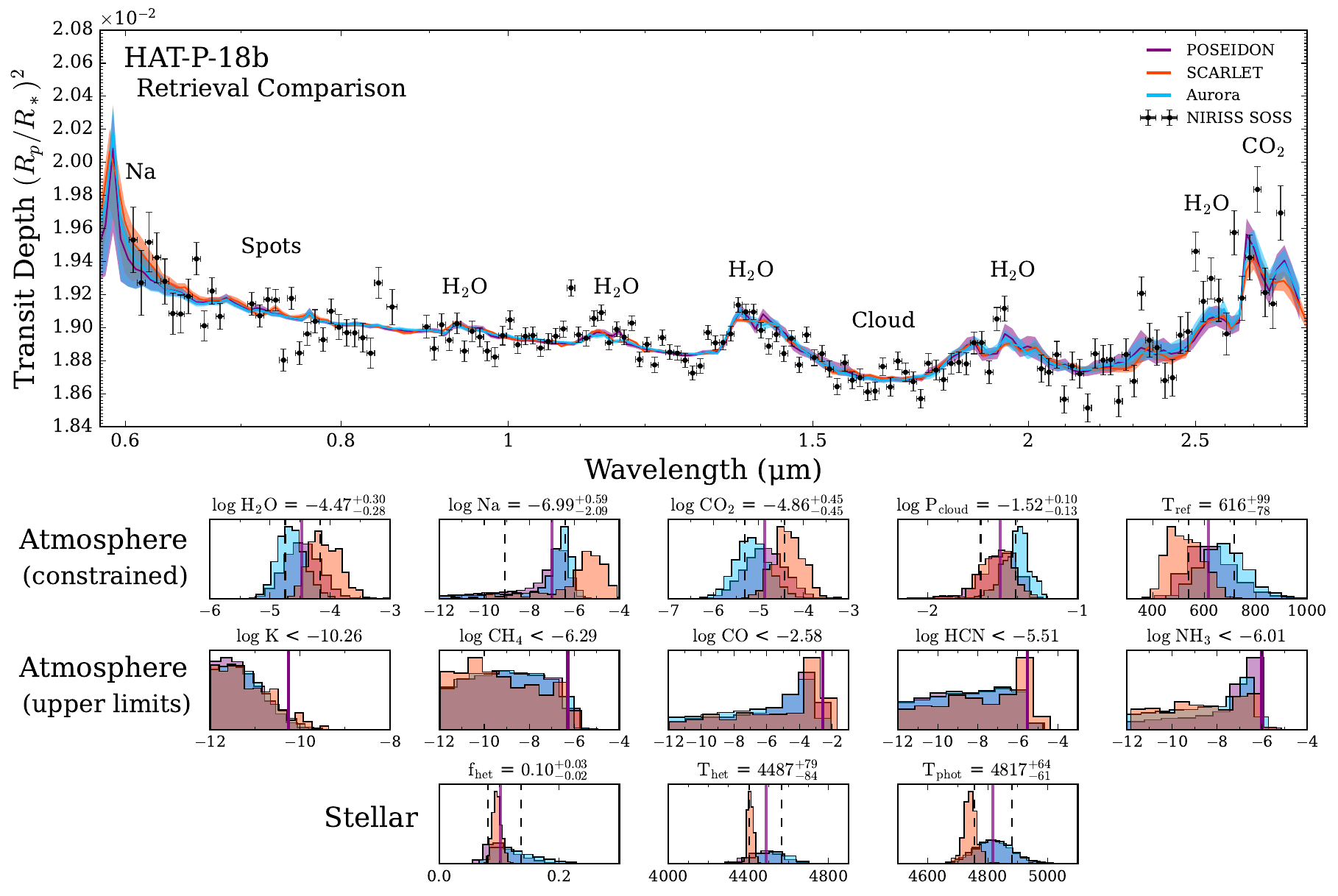}
    \caption{Atmospheric and stellar retrieval results for HAT-P-18\,b. \emph{Top}: Retrieved model transmission spectra from our JWST/NIRISS spectra (black errors). For each of the three retrieval codes (\textsc{Poseidon}; purple, \texttt{SCARLET}; orange, and \textsc{Aurora}; blue), the median retrieved spectrum (solid lines) and 1\,$\sigma$ confidence intervals (shaded contours) are shown. The most important model features required to explain HAT-P-18\,b's NIRISS transmission spectrum are annotated. All three codes adopt the \textit{one heterogeneity} model for this comparison. \emph{Bottom}: Posterior probability distributions corresponding to the retrieval model in the top panel. The top row shows retrieved atmospheric properties with constrained values, the middle row shows non-detected chemical species with abundance upper limits, and the bottom row shows the retrieved unocculted starspot properties. Each histogram is annotated with the retrieved median and $\pm$ 1\,$\sigma$ confidence intervals from the \textsc{Poseidon} retrieval for reference (see Table~\ref{tab:retrieval_results} for the results from all three codes).}
\label{fig:JWST_Retrieval_Summary}
\end{figure*}

Our retrieved atmospheric properties for HAT-P-18\,b are summarized in Figure~\ref{fig:JWST_Retrieval_Summary}. All three retrieval codes obtain retrieved transmission spectra consistent with the picture presented previously: HAT-P-18\,b's observed spectrum is shaped by absorption from H$_2$O and CO$_2$ (and likely Na), unocculted starspots, and a cloud deck. The posterior distributions in Figure~\ref{fig:JWST_Retrieval_Summary}, and corresponding confidence regions in Table~\ref{tab:retrieval_results}, provide our quantitative constraints on HAT-P-18\,b's atmospheric properties.

We obtain precise constraints on several atmospheric properties, including the H$_2$O and CO$_2$ abundances, cloud deck pressure, and atmospheric temperature. The retrieved H$_2$O abundance ($\log X_{\mathrm{H_2 O}} =  -4.47_{-0.28}^{+0.30}$ (\textsc{Poseidon}); $-4.11_{-0.22}^{+0.27}$ (\texttt{SCARLET}); $-4.65_{-0.22}^{+0.24}$ (\textsc{Aurora})) is $\sim$\,10$\times$ lower than the expected abundance for a solar-composition atmosphere under chemical equilibrium at HAT-P-18\,b's equilibrium temperature \citep{hartman2011}. Conversely, the retrieved CO$_2$ abundance ($\log X_{\mathrm{CO_2}} = -4.86_{-0.45}^{+0.45}$ (\textsc{Poseidon}); $-4.39_{-0.26}^{+0.38}$ (\texttt{SCARLET}); $-5.16_{-0.36}^{+0.40}$ (\textsc{Aurora})) is $>$ 100$\times$ higher than expected for a similar solar-composition atmosphere (cf. \citealt{woitke2018}). The retrieved Na abundance ($\log X_{\mathrm{Na}} = -6.99_{-2.09}^{+0.59}$ (\textsc{Poseidon}); $-5.45_{-1.59}^{+0.69}$ (\texttt{SCARLET}); $-6.65_{-1.68}^{+0.42}$ (\textsc{Aurora})) is more weakly constrained, being broadly consistent with a range of sub-solar to solar (\textsc{Poseidon} and \textsc{Aurora}) or even a super-solar composition (\texttt{SCARLET}). An opaque cloud deck, consistent with uniform terminator coverage ($f_{\mathrm{cloud}} > 0.83$, see Table~\ref{tab:retrieval_results}), is localized with a cloud top pressure of $\approx$ 30\,mbar ($\log P_{\mathrm{cloud}} = -1.52_{-0.13}^{+0.10}$ (\textsc{Poseidon}); $-1.53_{-0.13}^{+0.10}$ (\texttt{SCARLET}); $-1.40_{-0.10}^{+0.07}$ (\textsc{Aurora})). Finally, the retrieved P-T profile, over the pressure range probed by our NIRISS/SOSS observation, is isothermal at $\approx$ 600\,K ($T_{\mathrm{ref}} = 616_{-78}^{+99}$\,K (\textsc{Poseidon}); $523_{-65}^{+81}$\,K (\texttt{SCARLET}); $673_{-85}^{+90}$\,K (\textsc{Aurora})). We discuss the sensitivity of these inferences to the assumed stellar contamination model in Section~\ref{sec:results_stellar_contam_sensitivity}. We note that, while \texttt{SCARLET} favours slightly higher abundances and lower temperatures than \textsc{Poseidon} and \textsc{Aurora}, all three codes produce consistent retrieval results to within 1\,$\sigma$. The small differences between our retrieval results may be due to the slightly different model configuration and priors used by \texttt{SCARLET} (see Table~\ref{tab:retrieval_priors}).

\begin{table}
    \ra{1.2}
    \caption{Retrieval results for the \textit{one heterogeneity} model.}
    \begin{tabular*}{\columnwidth}{l@{\extracolsep{\fill}} lllll@{}}\toprule
    Parameter & \textsc{Poseidon} & \texttt{SCARLET} & \textsc{Aurora} \\ \midrule
    \textbf{Composition} & \\ 
    \hspace{0.5em} $\log X_{\mathrm{H_2 O}}$ & ${-4.47}_{-0.28}^{+0.30}$ & ${-4.11}_{-0.22}^{+0.27}$ & ${-4.65}_{-0.22}^{+0.24}$ \\ 
    \hspace{0.5em} $\log X_{\mathrm{Na}}$ & ${-6.99}_{-2.09}^{+0.59}$ & ${-5.45}_{-1.59}^{+0.69}$ & ${-6.65}_{-1.68}^{+0.42}$ \\ 
    \hspace{0.5em} $\log X_{\mathrm{CO_2}}$ & ${-4.86}_{-0.45}^{+0.45}$ & ${-4.39}_{-0.26}^{+0.38}$ & ${-5.16}_{-0.36}^{+0.40}$ \\ 
    \hspace{0.5em} $\log X_{\mathrm{K}}$ & $< -10.26$ & $< -9.87$ & $< -10.11$ \\
    \hspace{0.5em} $\log X_{\mathrm{CH_4}}$ & $< -6.29$ & $< -6.12$ & $< -6.30$  \\
    \hspace{0.5em} $\log X_{\mathrm{CO}}$ & $< -2.58$ & $< -2.26$ & $< -2.85$  \\
    \hspace{0.5em} $\log X_{\mathrm{HCN}}$ & $< -5.51$ & $< -5.01$ & $< -5.56$  \\
    \hspace{0.5em} $\log X_{\mathrm{NH_3}}$ & $< -6.01$ & $< -6.13$ & $< -6.21$  \\ \midrule
    \textbf{P-T profile} & \\ 
    \hspace{0.5em} $T_{\mathrm{ref}}$ & ${616}_{-78}^{+99}$ & ${523}_{-65}^{+81}$ & ${673}_{-85}^{+90}$ \\ \midrule
    \textbf{Aerosols} & \\
    \hspace{0.5em} $\log P_{\mathrm{cloud}}$ & ${-1.52}_{-0.13}^{+0.10}$ & ${-1.53}_{-0.13}^{+0.10}$ & ${-1.40}_{-0.10}^{+0.07}$ \\
    \hspace{0.5em} $f_{\mathrm{cloud}}$ & $> 0.87$ & $> 0.83$ & $> 0.90$ \\ \midrule
    \textbf{Stellar} & \\
    \hspace{0.5em} $f_{\mathrm{het}}$ & ${0.10}_{-0.02}^{+0.03}$ & ${0.10}_{-0.01}^{+0.01}$ & ${0.12}_{-0.03}^{+0.04}$ \\
    \hspace{0.5em} $T_{\mathrm{het}}$ & ${4487}_{-84}^{+79}$ & ${4408}_{-12}^{+16}$ & ${4513}_{-92}^{+85}$ \\
    \hspace{0.5em} $T_{\mathrm{phot}}$ & ${4817}_{-61}^{+64}$ & ${4740}_{-19}^{+22}$ & ${4821}_{-60}^{+62}$ \\
    \bottomrule
    \vspace{-0.5pt}
    \end{tabular*}
    \textit{Note:} We do not list results for unconstrained parameters (e.g., the other P-T profile parameters, since the retrieved profiles are essentially isothermal). `$<$' and `$>$' represent 2\,$\sigma$ upper and lower limits, respectively.
    \label{tab:retrieval_results}
\end{table}

We also obtain robust upper limits on the abundances of several important chemical species. Our retrievals strongly disfavour the presence of K ($\log X_{\mathrm{K}} \lesssim -10$ to 2\,$\sigma$), which could be due to condensation out of the gas phase. Similarly, we establish that CH$_4$ is at least $100\times$ less abundant than the expectation for a solar-composition atmosphere in chemical equilibrium ($\log X_{\mathrm{CH_4}} \lesssim -6$ to 2\,$\sigma$ vs. the expected $\log X_{\mathrm{CH_4}} \sim -4$; \citealt{woitke2018}), which could indicate photochemical dissociation \citep[e.g.,][]{moses2011,venot2012}. Although we see a tentative hint of CO in our posteriors (see Figure~\ref{fig:JWST_Retrieval_Summary}), additional observations at longer wavelengths, covering the 4.6\,$\mu$m CO band, would be required to robustly constrain the CO abundance (and hence the atmospheric C/O ratio). We also find upper limits on HCN and NH$_3$ ($\log X_{\mathrm{HCN}} \lesssim -5$; $\log X_{\mathrm{NH_3}} \lesssim -6$), which are consistent with both equilibrium and disequilibrium expectations for hot giant planets \citep[e.g.,][]{moses2011,venot2012}. These upper limits underline that SOSS data is of sufficiently high quality to provide scientifically meaningful constraints even on the abundances of non-detected chemical species.


\subsection{Unocculted Stellar Heterogeneities} \label{sec:results_unocculted_stellar_params}

We next turn to constraints on unocculted stellar heterogeneities. We first present results under the assumption of a single unocculted heterogeneity population with fixed $\log g$ --- a common approach in the literature --- before relaxing this assumption by considering more complex treatments of the TLSE. Finally, we examine the impact of different unocculted stellar heterogeneity models on the retrieved atmospheric properties of HAT-P-18\,b.

\begin{figure*}
    \centering
    \includegraphics[width=\textwidth]{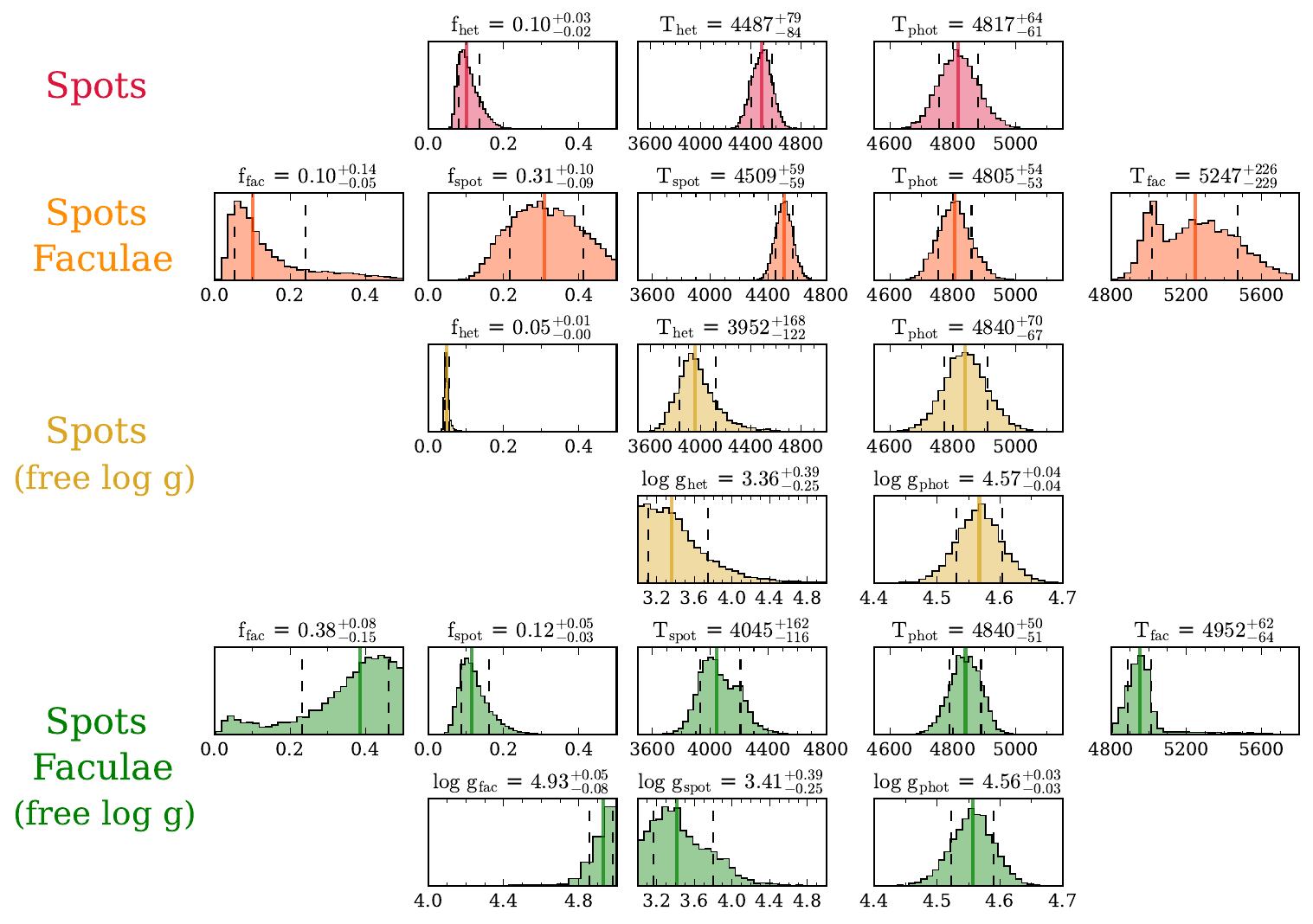}
    \caption{Unocculted stellar heterogeneity properties from the \textsc{Poseidon} retrieval analysis. Posterior distributions from four retrieval models are shown, from top to bottom: (i) a single population of stellar heterogeneities (spots) with $\log g_{\rm{spot}} = \log g_{\rm{phot}} = 4.57$, as in Figure~\ref{fig:JWST_Retrieval_Summary} (red); (ii) two distinct populations of spots and faculae, all with $\log g = 4.57$ (orange); (iii) same as the spot model, but with the $\log g$ of the spots and photosphere as free parameters (gold); and (iv) same as the spot + faculae model, but with free $\log g$ for all three stellar components (green). The histograms are ordered such that the columns (within each row) have the same physical interpretation between the different retrievals (e.g., $f_{\rm{het}}$ from the \textit{one heterogeneity} models corresponds to $f_{\rm{spot}}$ in the \textit{two heterogeneities} models). The retrieved atmospheric properties for HAT-P-18\,b are consistent across all four unocculted stellar heterogeneity models.}
\label{fig:JWST_unocculted_spot_param}
\end{figure*}

\subsubsection{One Heterogeneity Stellar Contamination Model} \label{sec:results_standard_TLS}

Assuming a single population of unocculted stellar active regions, our retrievals require unocculted starspots to explain HAT-P-18\,b's transmission spectrum (see Figure~\ref{fig:JWST_Retrieval_Summary} and Table~\ref{tab:retrieval_results}). The spots are $\approx$ 300\,K cooler than the photosphere ($T_{\mathrm{het}} = 4487_{-84}^{+79}$\,K (\textsc{Poseidon}); $4408_{-12}^{+16}$\,K (\texttt{SCARLET}); $4513_{-92}^{+85}$\,K (\textsc{Aurora}), compared to $T_{\mathrm{phot}} \approx 4800$\,K) and cover $\approx$ 10 \% of the stellar surface ($f_{\mathrm{het}} = 0.10_{-0.02}^{+0.03}$ (\textsc{Poseidon}); $0.10_{-0.01}^{+0.01}$ (\texttt{SCARLET}); $0.12_{-0.03}^{+0.04}$ (\textsc{Aurora})). We note that \texttt{SCARLET} has smaller uncertainties on the retrieved stellar parameters, which we attribute to the different parameterization of the stellar heterogeneity and hazes in the prior.

\subsubsection{More Complex Stellar Contamination Models} \label{sec:results_complex_TLS}

We find that the best-fitting stellar contamination model for HAT-P-18\,b is two distinct populations of unocculted heterogeneities --- spots and faculae --- with different surface gravities. We established this via a series of Bayesian model comparisons between the five stellar contamination models listed in Section~\ref{sec:TLS_models}. Our results, summarized in Table~\ref{tab:TLS_model_comparison}, demonstrate a significant improvement in the Bayesian evidence for the \textit{two heterogeneities with free surface gravity} model compared to the fiducial \textit{one heterogeneity} model we have focused on thus far (a Bayes factor of 77.5, equivalent to 3.4\,$\sigma$). We also find that the preference for unocculted heterogeneities vs. no stellar contamination is greater for the \textit{two heterogeneities with free surface gravity} model compared to the \textit{one heterogeneity} model (6.5\,$\sigma$ vs. 5.8\,$\sigma$). Nevertheless, we show below that adopting the preferred stellar contamination model does not significantly alter the atmospheric constraints for HAT-P-18\,b presented in Figure~\ref{fig:JWST_Retrieval_Summary}.

\begin{table}
    \ra{1.2}
    \caption{Unocculted stellar heterogeneity Bayesian model comparison.}
    \begin{tabular*}{\columnwidth}{l@{\extracolsep{\fill}} llllll@{}}\toprule
    Model & $\ln \mathcal{Z_{\rm{i}}}$ & $\mathcal{B}_{\rm{ref, i}}$ & $\ln \mathcal{B}_{\rm{ref, i}}$ & Det. Sig. \\ \midrule
    \textbf{\textit{Two het. (free log g)}} & 984.74 & Ref. & Ref & Ref. \\ 
    \textit{No stellar contam.} & 965.54 & 2.19\,$\times 10^8$ & 19.2 & 6.5\,$\sigma$ \\   
    \textit{One het.} & 980.39 & 77.5 & 4.35 & 3.4\,$\sigma$ \\ 
    \textit{One het. (free log g)} & 981.53 & 25.0 & 3.22 & 3.0\,$\sigma$ \\ 
    \textit{Two het.} & 983.57 & 3.24 & 1.18 & 2.1\,$\sigma$ \\ 
    \bottomrule
    \vspace{-0.5pt}
    \end{tabular*}
    \textit{Note:} $\ln \mathcal{Z_{\rm{i}}}$ is the Bayesian evidence of the $\mathrm{{i}^{th}}$ model. $\mathcal{B}_{\rm{ref, i}}$ is the Bayes factor of the reference model with respect to the $\mathrm{{i}^{th}}$ model. ``Det. Sig.'' indicates the detection significance, expressed in equivalent ``$\sigma$'' \citep[e.g.,][]{benneke2013}, for the reference model, highlighted in bold, vs. the $\mathrm{{i}^{th}}$ model.
    \label{tab:TLS_model_comparison}
\end{table}

We show in Figure~\ref{fig:JWST_unocculted_spot_param} that the choice of stellar contamination model can significantly alter inferences about unocculted stellar active regions. The most significant change is that retrievals with free $\log g$ favour much cooler spots ($\Delta T \approx -800$\,K vs. $\approx -300$\,K) than those with $\log g_{\mathrm{spot}} = \log g_{\mathrm{phot}}$. Indeed, the retrieved spot surface gravity is lower than the photosphere by $>$ 2\,$\sigma$. We similarly find that fitting for $\log g_{\mathrm{fac}}$ allows a more accurate determination of the faculae temperature compared to fixing the surface gravities. Overall, our preferred solution for HAT-P-18's unocculted active regions is as follows: spots and faculae cover $f_{\mathrm{spot}} = 12^{+5}_{-3}$\,\% and $f_{\mathrm{fac}} > 4$\,\% (2\,$\sigma$ lower limit) of HAT-P-18's surface, with temperatures of $T_{\mathrm{spot}} = 4045^{+162}_{-116}$\,K and $T_{\mathrm{fac}} = 4952^{+62}_{-64}$\,K and surface gravities of $\log g_{\mathrm{spot}} < 4.17$ (2\,$\sigma$ upper limit) and $\log g_{\mathrm{fac}} > 4.45$ (2\,$\sigma$ lower limit). These compare to a background photosphere with $T_{\mathrm{phot}} = 4840^{+50}_{-51}$\,K and $\log g_{\mathrm{phot}} = 4.56^{+0.03}_{-0.03}$. These results demonstrate that more complete stellar contamination models can provide deeper insights into the nature of unocculted stellar active regions.

\subsubsection{Sensitivity of Atmospheric Properties to Stellar Contamination Models} \label{sec:results_stellar_contam_sensitivity}

\begin{figure}
    \centering
    \includegraphics[width=\columnwidth]{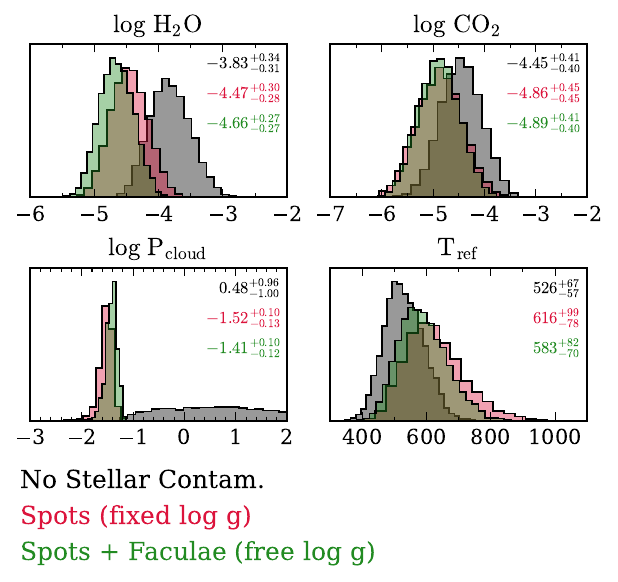}
    \caption{Impact of unocculted stellar contamination model on retrieved atmospheric properties. Posterior distributions are shown from three \textsc{Poseidon} retrieval models: \textit{no stellar contamination} (black), \textit{one heterogeneity} (red), and \textit{two heterogeneities with free surface gravity} (green). The \textit{one heterogeneity} (``spots'') model corresponds to the results in Figure~\ref{fig:JWST_Retrieval_Summary}. Without considering stellar contamination, one would retrieve a H$_2$O abundance biased 1--2\,$\sigma$ higher and find no cloud deck. The CO$_2$ abundance and atmospheric temperature are less sensitive to the stellar contamination model.}
\label{fig:atm_sensitivitity_to_stellar_contam}
\end{figure}

Finally, we find that, provided the spectroscopic imprint of unocculted heterogeneities is considered in the atmospheric model, the adopted stellar contamination model only minimally impacts the retrieved atmospheric properties for this data. Figure~\ref{fig:atm_sensitivitity_to_stellar_contam} shows that our preferred stellar contamination model (\textit{two heterogeneities with free surface gravity}) produces retrieved temperatures, cloud pressures, and H$_2$O and CO$_2$ abundances consistent with the simpler spot-only model with fixed surface gravity (\textit{one heterogeneity}). However, a retrieval not including stellar contamination would attribute the spectral slope (caused by spots and a cloud deck) to an atmospheric haze and erroneously retrieve a H$_2$O abundance 1--2\,$\sigma$ higher than when spots are included. Given that we do not see evidence of a haze when including a stellar heterogeneity model, we conclude that our NIRISS/SOSS data have sufficient wavelength coverage and precision to lift the degeneracy between an atmospheric haze power law and unocculted starspots. This mirrors a conclusion from \citet{rackham2023a}, where they demonstrated that haze-spot degeneracies can be lifted with sufficiently high-quality transmission spectra (see their Figure 22). Therefore, while the finer details of the stellar contamination model are less crucial if one only wishes to constrain the planetary atmosphere, a simple stellar contamination model (e.g., \textit{one heterogeneity}, as considered in studies such as \citealt{pinhas2018} and \citealt{rathcke2021}) is vital to obtain reliable atmospheric inferences for exoplanets transiting active stars.

\subsection{A Comparison between HST + Spitzer and JWST}\label{sec:results_Hubble}

We additionally ran retrievals on our newly reduced HST and \textit{Spitzer} observations to contextualize our JWST results. We consider two \textsc{Poseidon} retrievals, a model without stellar contamination and with a single unocculted stellar heterogeneity, adopting the same retrieval configuration and priors as used in the JWST analysis (see section~\ref{sec:retrieval_config} and Table~\ref{tab:retrieval_priors}). The only difference compared to our previous approach is an altered model wavelength grid (0.9--5.3\,$\mu$m at $R$ = 20,000) to cover the \textit{Spitzer} data at longer infrared wavelengths.

\begin{figure}
    \centering
    \includegraphics[width=\columnwidth]{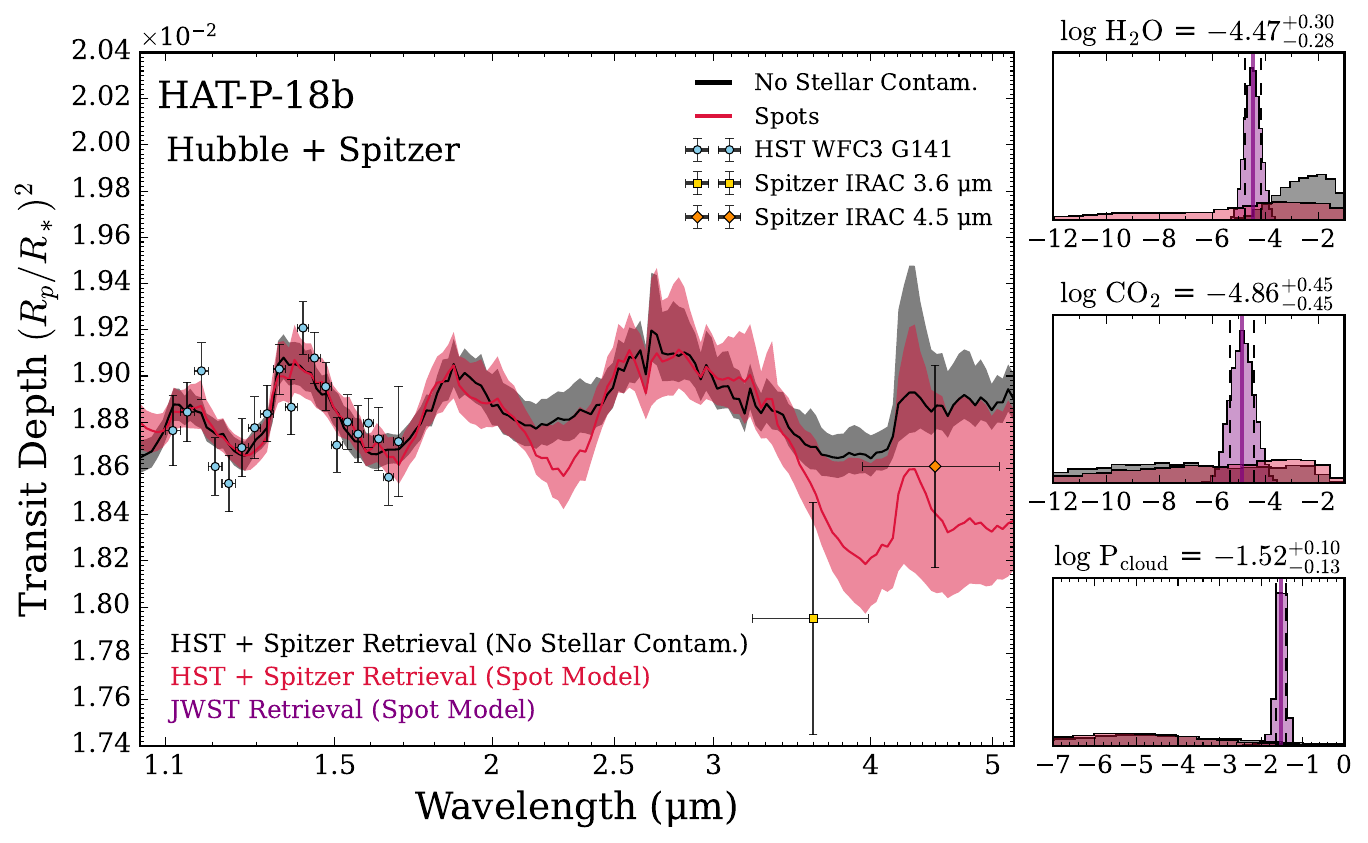}
    \caption{Atmospheric retrieval of Hubble and \textit{Spitzer} transmission spectra of HAT-P-18\,b. \emph{Left}: Retrieved spectrum assuming no stellar contamination (black) and with a single heterogeneity (red), shown via their median (solid lines) and 1\,$\sigma$ confidence intervals. \emph{Right}: Corresponding posterior probability distributions. The JWST NIRISS/SOSS retrieval results from the single heterogeneity \textsc{Poseidon} retrieval (see Figure~\ref{fig:JWST_Retrieval_Summary}) are overlaid, and the retrieved median and $\pm$ 1\,$\sigma$ parameter constraints annotated, demonstrating that NIRISS/SOSS data provides significantly improved constraints on the atmospheric composition and cloud pressure compared to Hubble and \textit{Spitzer} data.}
\label{fig:HST_Spitzer_retrieved_spectrum}
\end{figure}

Figure~\ref{fig:HST_Spitzer_retrieved_spectrum} compares our retrieved spectrum and atmospheric constraints from HST and \textit{Spitzer} to our previously presented JWST results. We focus on the abundances of H$_2$O, CO$_2$, and the cloud pressure for this comparison since these are the most well-constrained model parameters from our JWST/NIRISS data (see section~\ref{sec:results_atmo}). Both retrieval models infer an attenuated absorption feature in the HST/WFC3 data, suggesting a degree of cloud opacity ($\log P_{\mathrm{cloud}} \lesssim -2$ to 1$\sigma$). The model without stellar contamination attributes the absorption near 1.1 and 1.4\,$\mu$m to H$_2$O, with an abundance constraint at the order-of-magnitude level ($\log X_{\mathrm{H_2 O}} =  -2.43_{-1.05}^{+0.87}$). However, the unocculted starspot model, which obtains an improved fit to the \textit{Spitzer} data, finds the HST/WFC3 data can be explained by either H$_2$O absorption or CH$_4$ in combination with starspots. This three-way degeneracy between $f_{\rm{het}}$, $\log X_{\mathrm{H_2 O}}$, and $\log X_{\mathrm{CH_4}}$ --- caused by the lack of data shortwards of 1.0\,$\mu$m --- removes the H$_2$O detection one could claim under the assumption of no stellar contamination. Therefore, HAT-P-18\,b's H$_2$O abundance is essentially unconstrained from the HST and \textit{Spitzer} observations. In contrast, our JWST/NIRISS observation has the wavelength coverage to break the degeneracies between unocculted starspots and the atmospheric composition, affording detailed characterization of HAT-P-18\,b's atmosphere.

\section{Summary \& Discussion}\label{sec:discussion}

This work has led to important inferences on both the stellar activity of HAT-P-18 and the atmosphere of its hot Saturn. Our main results are as follows:

\begin{itemize}
    \item The transit light curves show evidence of a spot-crossing event ($>$\,5\,$\sigma$); we inferred that the most likely parameters for that occulted spot are a position on the projected stellar surface of (x, y) = (0.090 $\pm$ 0.005, 0.42 $\pm$ 0.05)\,R\textsubscript{*}, with a radius of 0.116 $\pm$ 0.014\,R\textsubscript{*}, and a temperature colder than the star of $\Delta T$ = -93 $\pm$ 15 K.
    \item The main features in the transmission spectrum retrieved from NIRISS/SOSS observation are multiple absorption features produced by water vapour (12.5\,$\sigma$) with a retrieved abundance of $\log$ H$_2$O $\approx -4.4 \pm 0.3$, a rise at redder wavelengths due to a CO$_2$ absorption feature (7.3\,$\sigma$) and evidence of Na (2.7\,$\sigma$). Also, there is a slope towards bluer wavelengths caused by unocculted starspots (5.8\,$\sigma$) and a uniform, grey cloud deck (7.4\,$\sigma$) that mutes some absorption features.
    \item Modelling stellar heterogeneities led to four slightly different solutions for the occulted spot, and we showed that different treatments of the unocculted active regions could be used. Fortunately, we found that the different solutions for the active regions had no significant impact on the retrieved transmission spectrum and atmospheric properties of HAT-P-18\,b.
    \item Modelling spot spectra is best achieved with stellar models with lower surface gravities than the stellar photosphere. For the most likely solution of the occulted spot, that is a $\Delta \log g$ = 1.16 $\pm$ 0.19\,dex, which is in perfect agreement with our solutions for the unocculted spots with free $\log g$.  
\end{itemize}

We proceed to discuss the implications of our results and their potential impact on exoplanetary atmosphere studies in transmission spectroscopy.

\subsection{The Atmosphere of HAT-P-18 b in Context} \label{sec:discussion_atmosphere}

The best-fitting atmosphere model confirms the presence of H\textsubscript{2}O and clouds in the terminator region of HAT-P-18\,b, which is in agreement with previous work \citep{tsiaras2018}. As in \citet{fu2022}, our transmission spectrum shows a slope towards bluer wavelengths; however, to explain it, instead of unocculted stellar heterogeneities, these authors inferred Rayleigh haze scattering, as also suggested by \citet{kirk2017} from a ground-based observation. The TLSE, which was not taken into consideration in those studies, is likely responsible for that difference, as well as the 10$\times$ higher water abundance in \citet{fu2022}. We also find 10$\times$ less CO\textsubscript{2} than was reported by \citet{fu2022}, and we furthermore cannot confirm their detection of CH\textsubscript{4}; both differences could come from the data reduction and modelling differences. 

\begin{figure}
    \centering
    \includegraphics[width=\columnwidth]{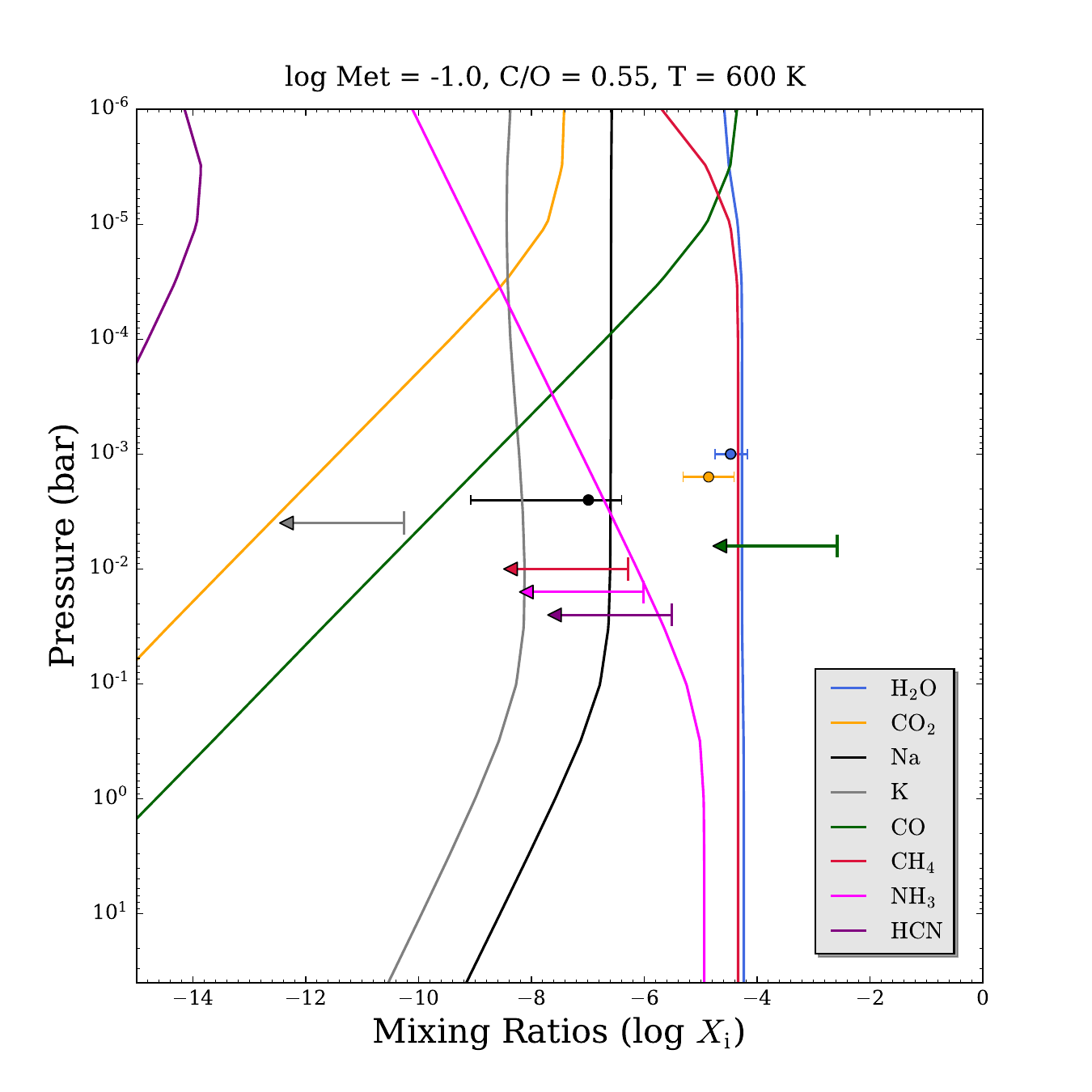}
    \caption{Comparison between an equilibrium chemistry model (coloured curves) and the retrieved mixing ratios from the \textsc{Poseidon} \textit{one heterogeneity} model (error bars and arrows). The error bars correspond to 1$\sigma$ constraints for the inferred chemical species (H$_2$O, CO$_2$, and Na), whilst the arrows correspond to 2$\sigma$ upper limits for the non-detected species. The \texttt{FastChem 2} equilibrium model shown assumes a solar C/O ratio and 10$\times$ sub-solar metallicity (chosen based on the retrieved H$_2$O mixing ratio) and a temperature consistent with the retrieved terminator temperature (600\,K).}
\label{fig:Eq_chem_vs_retrieved}
\end{figure}

We show in Figure \ref{fig:Eq_chem_vs_retrieved} our retrieved chemical abundances in comparison to an equilibrium chemistry model from \texttt{FastChem 2} \citep{stock2022} at 600 K with a metallicity of 0.1\,dex and a solar C/O ratio. We selected 10$\times$ sub-solar metallicity for this reference equilibrium model based on our retrieved sub-solar H$_2$O abundance (which is our most precisely constrained abundance), allowing us to compare the other retrieved molecular abundances to equilibrium expectations. The retrieved Na abundance is consistent with this equilibrium model, as are the non-detections of CO, HCN and NH$_3$. On the other hand, our retrieved abundance of CO\textsubscript{2} is significantly more abundant, whereas CH\textsubscript{4} and K are depleted by at least two orders of magnitude. We also computed equilibrium chemistry models with different C/O ratios, but the abundances did not change significantly. The differences between our free-retrieval results and the expectations from equilibrium chemistry may be indicative of either 1) non-equilibrium effects, like photochemistry for CH\textsubscript{4} and clouds for K or 2) a demonstration of the sensitivity of free-retrievals to individual points which could be driving an anomalously high CO$_2$ abundance. Methane, which would be expected at the temperature of HAT-P-18\,b, was also not detected in the atmosphere of the slightly warmer hot-Saturn WASP-39\,b \citep{feinstein2022}. The C/O ratio is still uncertain for HAT-P-18\,b's atmosphere and the need for longer wavelength observations with other instruments, such as NIRSpec G395H, are required to measure all the main carbon-bearing species and further investigate possible CH$_4$ signatures in the infrared.

Our detection of CO\textsubscript{2} may be spurious, and the abundance estimate unreliable. An edge effect in the NIRISS observation may be the culprit, as the CO\textsubscript{2} band is right at the red end of the SOSS spectrum, even exceeding it slightly. Longer wavelength data with the Near Infrared Spectrograph (NIRSpec) G395H grating would allow the detection of CO\textsubscript{2} to be confirmed and its abundance refined.

Our atmospheric analyses consistently find a uniform cloud deck as the only aerosol required to explain HAT-P-18\,b's NIRISS/SOSS transmission spectrum. Atmospheric studies of hot Jupiters often predict a degree of patchy clouds in hot Jupiter atmospheres \citep[e.g.,][]{parmentier2016,komacek2022}. However, despite allowing for patchy clouds, our retrievals all favor terminator cloud fractions $> 83\%$ (to 2$\sigma$). Future general circulation model studies focused on HAT-P-18\,b can further investigate whether uniform cloud coverage is more favoured for this Saturn-mass planet with a colder equilibrium temperature ($\sim$ 850\,K) than many hot Jupiters. Additional observations in the infrared will also be informative.

\subsection{Legacy of Hubble \& \textit{Spitzer} in Light of JWST Results}
For our HST and \textit{Spitzer} transmission spectrum, the model without stellar contamination finds an abundance of H\textsubscript{2}O (log H\textsubscript{2}O =  $-2.43_{-1.05}^{+0.87}$) and a cloud pressure ($\log P_{\mathrm{cloud}} \lesssim -2$ $\approx$ 10\,mbar) in good agreement with the findings of \citet{tsiaras2018} (log H\textsubscript{2}O = -2.63 $\pm$ 1.18 and $\log P_{\mathrm{cloud}}$ = -2.18 $\pm$ 0.91 $\approx$ 6.6\,mbar). However, we showed that a model considering the stellar contamination with a single heterogeneity removes the H\textsubscript{2}O detection because of a three-way degeneracy between the coverage fraction of heterogeneities and the abundances of H\textsubscript{2}O and CH\textsubscript{4} due to the lack of the visible spectrum with HST/WFC3. 

The impact of spots on optical transmission spectra has been known for some time to mimic a Rayleigh scattering slope (e.g., \citealp{mccullough2014,rackham2018}). Also, for exoplanets orbiting red dwarf stars, the 1.4\,$\mu$m water absorption feature observed with HST/WFC3 was sometimes debated to come from potential water molecules in their cold starspots (e.g., \citealp{barclay2021}). Now, with NIRISS/SOSS, we can break the degeneracy between a hazy atmosphere and starspots, and potentially for colder stars, we could confirm if water absorption features come from starspots or the exoplanetary atmosphere.

HST and \textit{Spitzer} inference studies have shown that water vapour, alkali metals, as well as clouds are common in hot giant planets (e.g., \citealp{madhusudhan2019}). Still, there often remained a degeneracy between water and clouds, preventing robustly retrieving the abundance of H\textsubscript{2}O and the location of the cloud deck (e.g., \citealp{benneke2012,benneke2013}). In this work, we confirmed these detections of water vapour, clouds and potentially sodium in a hot giant atmosphere, and we demonstrate that the improved wavelength coverage provided by NIRISS/SOSS is sufficient to break the cloud-metallicity degeneracy. 

\subsection{Challenges of Occulted Spot Analysis in the Era of JWST}

Our work adds to the growing body of literature demonstrating that the inference of the transit and occulted spots properties with JWST observations can and should be done to avoid possible biases in atmospheric estimates (e.g., \citealp{bixel2019,rackham2023a}). A semi-analytic tool like \texttt{spotrod} \citep{beky2014}, used here, is less computationally expensive compared to other tools that use a pixelation approach to model the stellar disc. Parallelization of those tools could be a simple solution to further improve the computation time. 

Moreover, we found that modelling active features from light curves, with the current tools and knowledge, is a degenerate problem. The standard practice of fixing the orbital parameters using a white light curve fit, which also fixes the occulted spot size and position, may prevent ruling out some families of solutions. Therefore, we are currently developing a package to simultaneously fit all spectral channels using both wavelength-dependent and wavelength-independent parameters. Still, theoretical advances are needed to understand the limits of inferences from occultations and break the degeneracy between starspot size and temperature. Fortunately, those degeneracies do not impact the retrieved transmission spectrum significantly. Furthermore, \texttt{spotrod} assumes that active features have a circular shape and the same limb-darkening law as the star. This can also impact the retrieved properties and make it more difficult to lift some degeneracies. The occulted feature may, moreover, have a more complex structure consisting of a cool spot and a hot facula, which could explain the smaller temperature difference retrieved for the occulted spot compared to the unocculted spots. Nonetheless, the retrieved stellar fraction of the occulted spot $f_{\mathrm{oc.\,spot}} = 1.35 \pm 0.02\,\%$ is coherent with our preferred solution for the coverage fraction of unocculted spots $f_{\mathrm{unoc.\,spot}} = 12^{+5}_{-3}$\,\%. 

\subsection{Accounting for Unocculted Heterogeneities in JWST Transmission Spectra}
Our best-fitting model suggests that the slope towards bluer wavelengths in this NIRISS/SOSS observation comes mainly from unocculted stellar heterogeneities. Our work shows that we should constrain planetary atmosphere and stellar contamination properties simultaneously with JWST NIRISS/SOSS data; otherwise, there is a risk of biasing atmospheric inferences. However, we could not distinguish clearly which of the four stellar contamination treatments investigated (one and two heterogeneities, one and two heterogeneities with free surface gravity) is the most plausible, but we were able to show that a particular choice of treatment does not significantly impact the atmospheric inferences.

Furthermore, the spectra of active regions are currently modelled with 1D stellar models, which is another limitation, particularly for faculae \citep{norris2017,johnson2021}, because they fail to capture aspects of facular contrasts, such as the center-to-limb variation (e.g., \citealp{yeo2013}) and the dependence on stellar metallicity \citep{witzke2018}. More theoretical work, such as magnetohydrodynamic simulations of magnetic features, is needed to understand better and, thus, model the contrast of active regions more accurately as a function of stellar fundamental parameters and activity. In the meantime, we suggest that spots' flux spectra be modelled with stellar models with lower surface gravities than the star.

\section{Conclusion}\label{sec:conclu}
We presented atmospheric retrieval analyses of two transmission spectra of HAT-P-18\,b obtained with JWST NIRISS/SOSS, and HST/WFC3 with \textit{Spitzer}/IRAC. We also modelled a spot-crossing event in the transit observed with NIRISS/SOSS. We confirmed that the wavelength coverage and spectral resolution of NIRISS/SOSS considerably improve the detection significance and the abundance constraints on chemical species compared to observations from HST and \textit{Spitzer} combined. In including stellar heterogeneities in transit fits and atmospheric retrievals, we implemented new model considerations designed to fit the local surface gravity of stellar heterogeneities. This work is instructive for the upcoming JWST transit observations, particularly by informing the community on disentangling stellar and planetary atmosphere signals. 

\section*{Acknowledgements}
We thank Steven Rogowski for his initial work on HST and \textit{Spitzer} data. This work is based on observations made with JWST, as well as the Hubble and Spitzer Space Telescopes. This project is undertaken with the financial support of the Canadian Space Agency. M.F.T. acknowledges financial support from the Fonds de Recherche du Québec — Nature et technologies (FRQNT) and funding from the Trottier Family Foundation in their support of iREx. R.J.M. acknowledges support for this work provided by NASA through the NASA Hubble Fellowship grant HST-HF2-51513.001, awarded by the Space Telescope Science Institute, which is operated by the Association of Universities for Research in Astronomy, Inc., for NASA, under contract NAS5-26555. M.R. acknowledges financial support from the Natural Sciences and Engineering Research Council of Canada (NSERC) as well as from FRQNT and iREx. C.P. acknowledges support from FRQNT and TEPS Ph.D. scholarships and the NSERC Vanier scholarship. K.M. acknowledges financial support from the FRQNT. O.L. acknowledges financial support from FRQNT and iREx. D.J. is supported by NRC Canada and by an NSERC Discovery Grant. J.D.T was supported for this work by NASA through the NASA Hubble Fellowship grant HST-HF2-51495.001-A awarded by the Space Telescope Science Institute, which is operated by the Association of Universities for Research in Astronomy, Incorporated, under NASA contract NAS5-26555.
\section*{Software}
\begin{itemize}
    \renewcommand\labelitemi{--}
    \item \texttt{astropy}; \citet{astropy2013,astropy2018}
    \item \texttt{batman}; \citet{kreidberg2015}
    \item \texttt{dynesty}; \citet{speagle2020}
    \item \texttt{emcee}; \citet{foreman-mackey2013}
    \item \texttt{ExoTiC-LD}; \citet{wakeford2022}
    \item \texttt{ipython}; \citet{perez2007}
    \item \texttt{juliet}; \citet{espinoza2019b}
    \item \texttt{matplotlib}; \citet{hunter2007}
    \item \texttt{nestle}; \citet{skilling2004}
    \item \texttt{numpy}; \citet{harris2020}
    \item \textsc{Poseidon}; \citet{macdonald2023}
    \item \texttt{scipy}; \citet{virtanen2020}
    \item \texttt{spotrod}; \citet{beky2014} 
    \item \texttt{supreme-SPOON}; \citet{radica2023}
\end{itemize}

\section*{Data Availability}
All data used in this study are publicly available from the Barbara A. Mikulski Archive for Space Telescopes\footnote{\url{https://mast.stsci.edu/portal/Mashup/Clients/Mast/Portal.html}} and the Spitzer Heritage Archive\footnote{\url{https://sha.ipac.caltech.edu/applications/Spitzer/SHA/}}.



\bibliographystyle{mnras}
\bibliography{hatp18b.bib} 




\appendix
\section{Additional Reduction} \label{sec:ind_pipeline}

We carried out an independent reduction on the HAT-P-18\,b SOSS TSO using \texttt{NAMELESS} (described in depth in \citealp{coulombe2022}). We applied all stage 1 steps of the \texttt{jwst} pipeline except for the dark current subtraction. We subsequently applied the world coordinate system, source type, and flat field steps of stage 2. We proceeded with the background subtraction, scaling independently the two regions separated by the jump of the model background provided by STScI, as described in detail in \citet{lim2023}. Treatment of the 1/$f$ noise was performed by scaling each individual column of the trace independently and finding the value that results in the lowest Chi-square for a given column and integration \citep{coulombe2022}. Finally, we extracted the 2D spectra using the \texttt{getSimpleSpectrum} function of the \texttt{transitspectroscopy} pipeline\footnote{\url{https://github.com/nespinoza/transitspectroscopy}} with an aperture width of 30 pixels.

\begin{figure*}
	\centering
	\includegraphics[width=\textwidth]{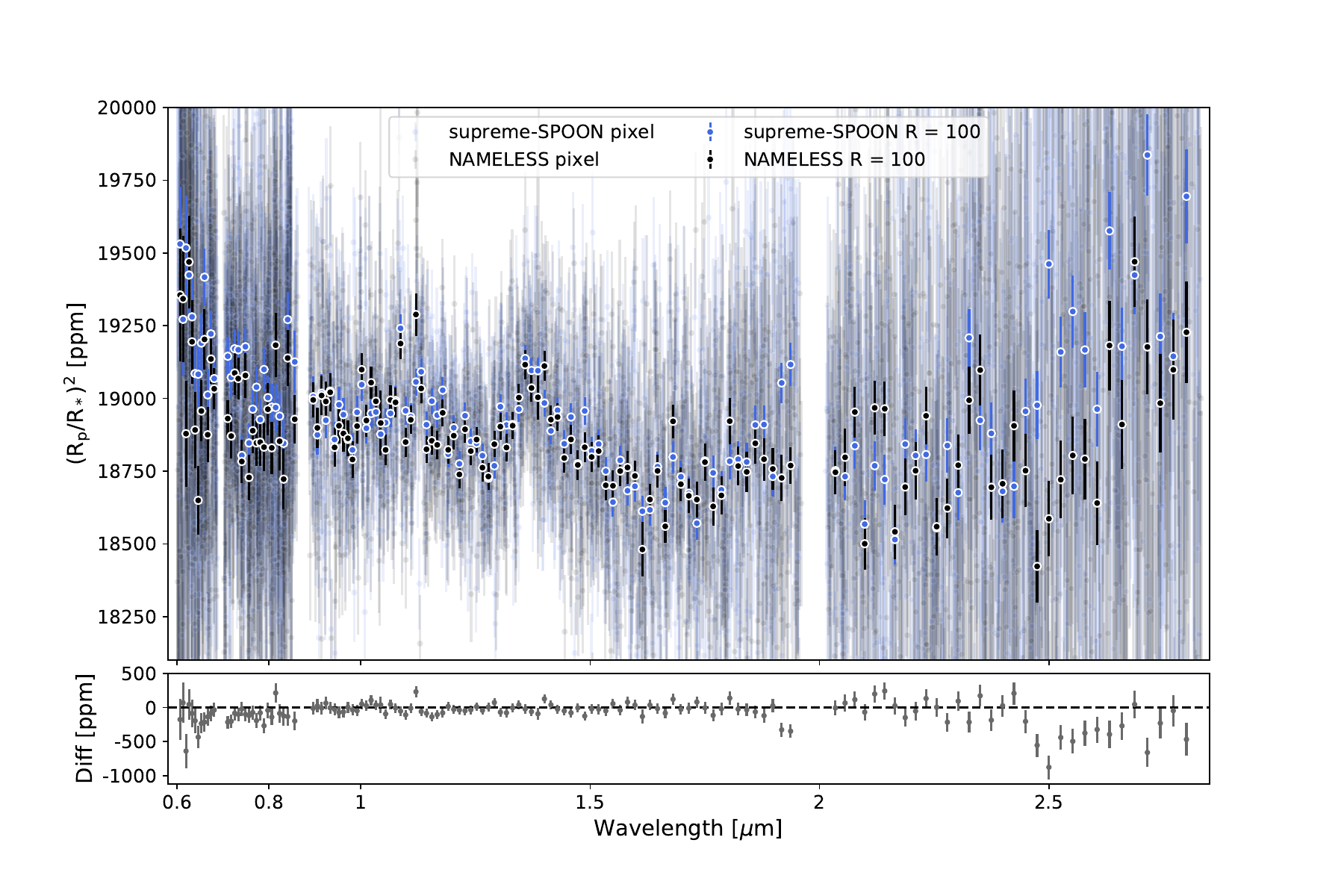}
    \caption{Comparison of transmission spectra for HAT-P-18\,b obtained with two different pipelines: \texttt{supreme-SPOON} (blue) and \texttt{NAMELESS} (black). \emph{Top}: Transmission spectra are shown binned to a constant resolving power of $R$ = 100 (darker points) and at the pixel resolution (faded points). \emph{Bottom}: Difference between the \texttt{NAMELESS}' and \texttt{supreme-SPOON}'s pipeline.
    \label{fig:ts_comparison}}
\end{figure*}

We perform the light curve fitting on the extracted spectrophotometric observation in an identical manner to the one described in Section \ref{sec:lightcurve}. We fix the orbital parameters ($T_\textrm{0}$, $b$, $a/R_\textrm{*}$) to the best-fitting values from the order 1 white light curve fit of the \texttt{supreme-SPOON} pipeline to ensure consistency. A comparison of the best-fitting white light curve parameters for NIRISS/SOSS with the two different pipelines is available in Table \ref{tab: WLC Parameters}. We note that the parameter values of $R_\textrm{p}/R_\textrm{*}$, \textit{b} and $a/R_\textrm{*}$ differ by 1.2-1.3\,$\sigma$. These parameters are correlated (e.g., see the corner plot for one of the broadband light curve fits; Figure \ref{fig:corner_spot}), and each reduction may simply prefer a slightly different place in the same family of solutions with a comparable agreement. The retrieved transmission spectrum, along with the one from the reference \texttt{supreme-SPOON} reduction, are shown in Figure \ref{fig:ts_comparison}. These transmission spectra obtained through two different pipelines are in overall good agreement; showing consistent transit depths and features. However, the amplitude of the CO\textsubscript{2} and Na features are different between the two reductions \citep[similarly to, e.g.,][]{radica2023} and would lead to some differences in the retrieved abundances.
\begin{table*}
\caption{Comparison of best-fitting ``white'' light curve parameters}
\label{tab: WLC Parameters}
    \begin{tabular}{cccccccc}
        \toprule
        Pipeline &Spot-crossing& $\rm t_0$ [BJD - 2400000] & $\rm R_p/R_*$ & $\rm b$ & $\rm a/R_*$ & q$_1$ & q$_2$ \\
        \midrule
        \textbf{NIRISS}\\ 
        \texttt{supreme-} & Masked&59743.353393$^{+0.000018}_{-0.000017}$  & 0.1377$^{+0.0003}_{-0.0003}$ & 0.398$^{+0.009}_{-0.009}$ & 15.31$^{+0.05}_{-0.05}$ & 0.20$^{+0.02}_{-0.02}$ & 0.31$^{+ 0.05}_{-0.05}$ \\
        \texttt{SPOON}& Modelled&59743.353403$^{+0.000019}_{-0.000017}$  & 0.1379$^{+0.0003}_{-0.0004}$ & 0.429$^{+0.010}_{-0.010}$ & 15.32$^{+0.05}_{-0.05}$ & 0.25$^{+0.03}_{-0.02}$ & 0.40$^{+ 0.05}_{-0.05}$\\\\
        \texttt{NAMELESS} &Masked& 59743.35340$^{+0.00002}_{-0.00002}$  & 0.1371$^{+0.0004}_{-0.0004}$ & 0.380$^{+0.013}_{-0.015}$ & 15.42$^{+0.07}_{-0.07}$ &  0.19$^{+0.02}_{-0.02}$ &  0.33$^{+ 0.06}_{-0.05}$\\ \\ \midrule
        \textbf{WFC3} &---& 57430 & 0.1373$^{+0.0004}_{-0.0004}$ & 0.373$^{+0.019}_{-0.019}$ & 16.69$^{+0.15}_{-0.15}$ & 0.179179 &0.296171\\\bottomrule \\
    \multicolumn{8}{l}{\footnotesize \textit{Note:} For the HST/WFC3 white light curve fit, the mid-transit time and the two parameters of the quadratic limb darkening law were} \\
    \multicolumn{8}{l}{\footnotesize fixed to those values. Note that the impact parameter value for the fit with the spot-crossing modelled is slightly higher due to a diffe-}\\
    \multicolumn{8}{l}{\footnotesize rent model being used (\texttt{spotrod} instead of \texttt{batman}). There are value differences for the limb darkening coefficients because of the}\\
    \multicolumn{8}{l}{\footnotesize different bandpasses (see text for exact wavelength ranges).}\\
    \end{tabular}
\end{table*}

\section{Additional Materials}
\begin{figure*}
	\centering
	\includegraphics[width=\linewidth]{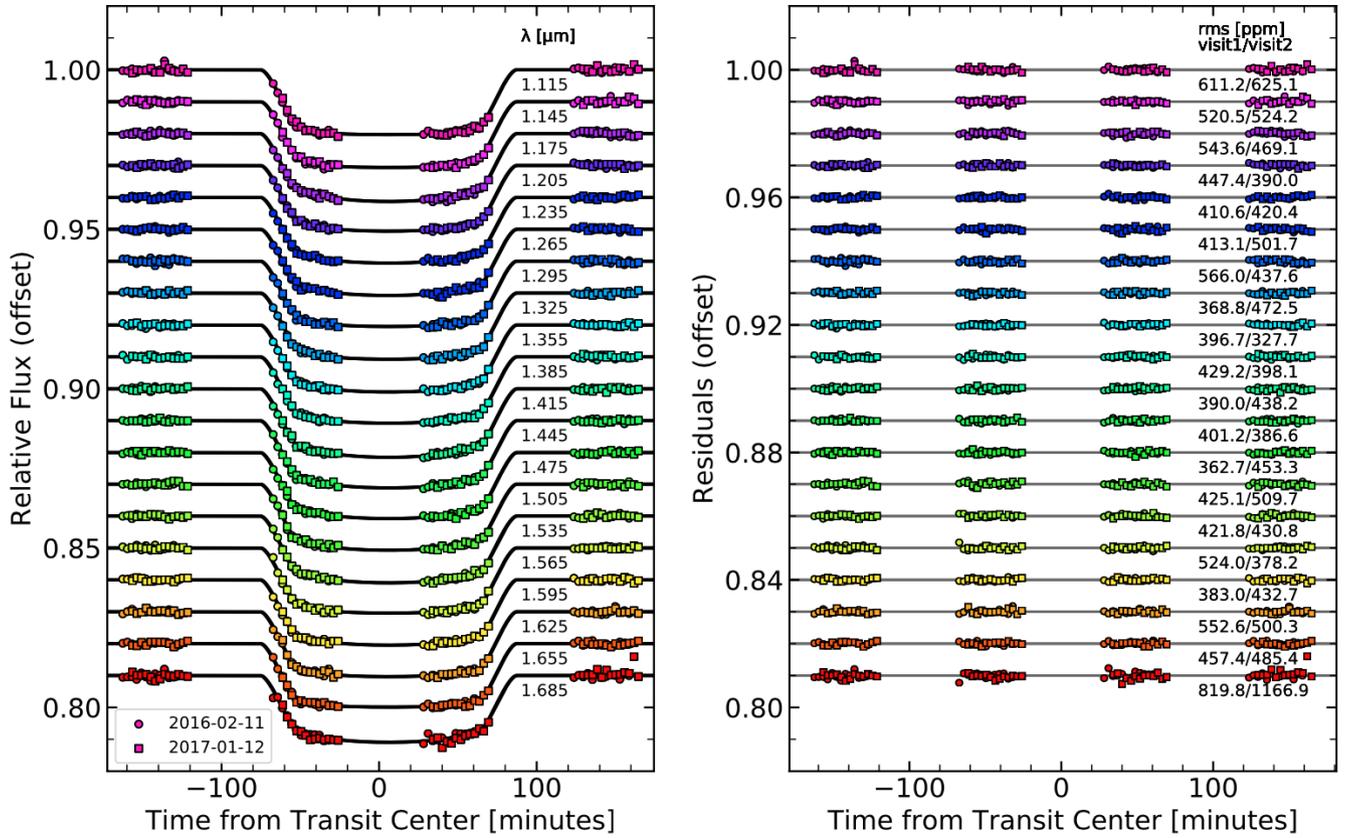}
    \caption{HST/WFC3 binned spectrophotometric light curves, along with the best-fitting transit models overplotted (black) for both visits. Light curves are arbitrarily offset for clarity. \emph{Right}: Residuals from the different light curve fits with the associated root-mean-square (RMS) scatter.
    \label{fig:HST lc}}
\end{figure*}

\begin{figure*}
    \centering
    \includegraphics[width=0.8\textwidth]{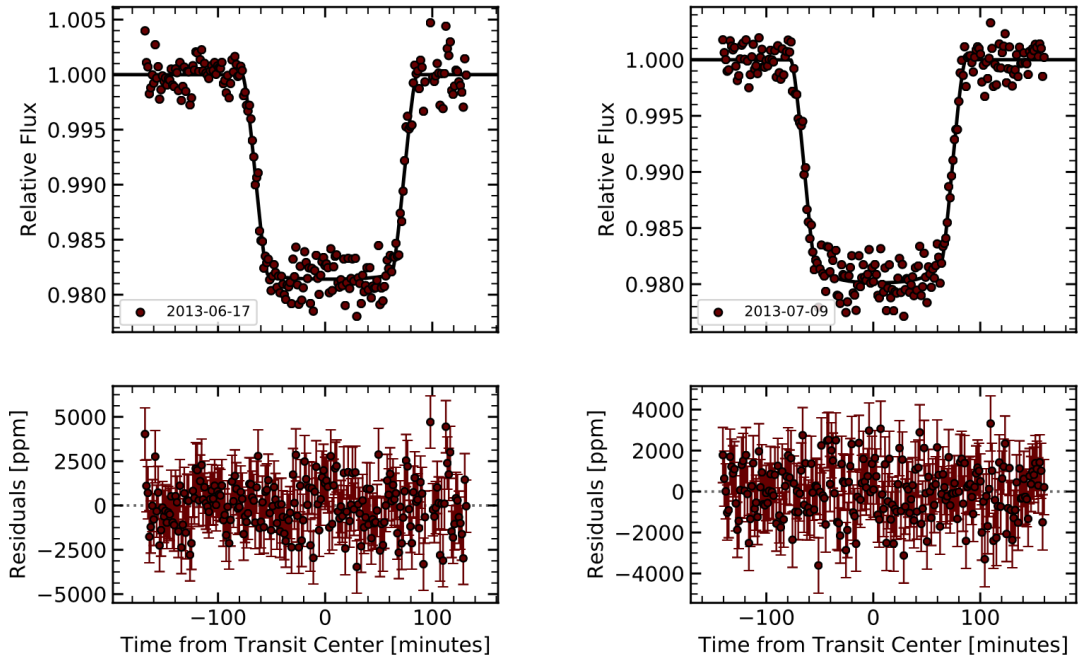}
    \caption{\textit{Spitzer}/IRAC light curves. \emph{Top}: Best-fitting transit models (black), overlaid with the systematics-corrected data (red circles), showing channel 1 (3.6\,$\mu$m) and channel 2 (4.5\,$\mu$m) in the left and right panel, respectively. \emph{Bottom}: Residuals from the light curve fits. 
    \label{fig:Spitzer lc}}
\end{figure*}
 \begin{figure*}
    \centering
    \includegraphics[width=0.7\linewidth]{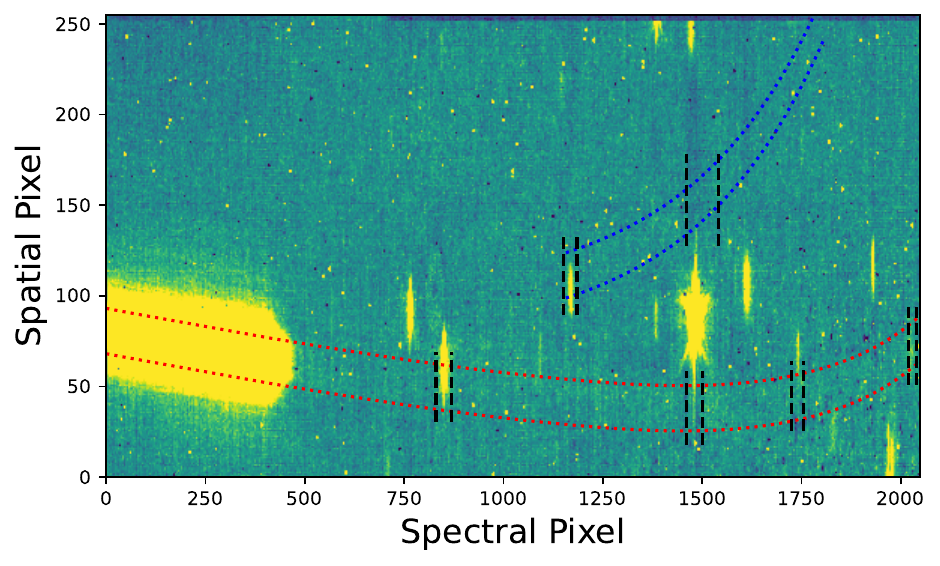}
    \caption{Exposure in the NIRISS/SOSS F277W filter ($2.5\leq\lambda\leq2.85$), highlighting the positions of undispersed contaminants. The extraction apertures for orders 1 and 2 are denoted in red and blue respectively. Regions of each order that are masked due to the presence of a contaminant are denoted by the black dashed lines (see text for exact wavelength ranges that are masked).}
    \label{fig:f277w}
\end{figure*}
 \begin{figure*}
	\centering
	\includegraphics[width=\textwidth]{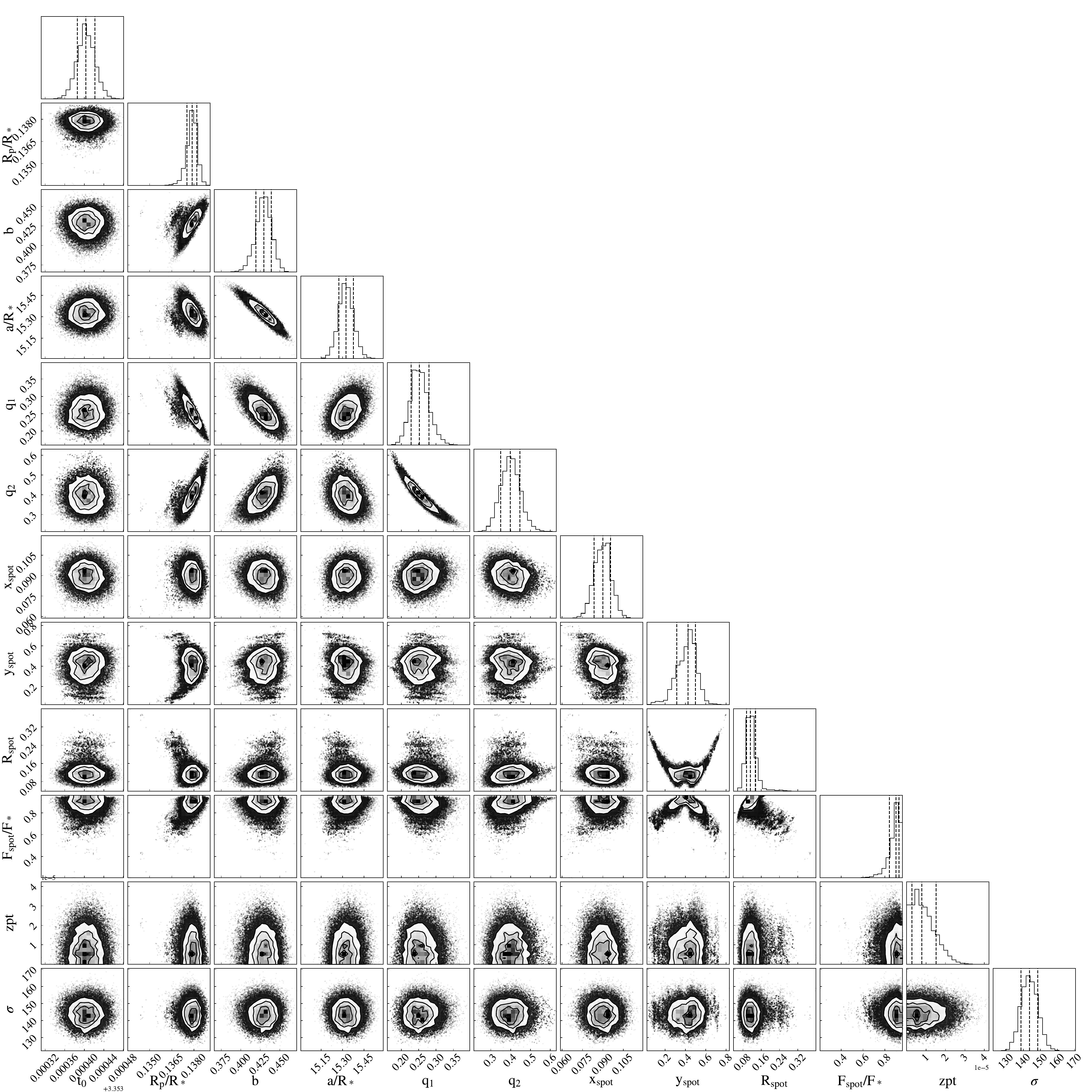}
    \caption{Posterior probability distributions from the NIRISS/SOSS broadband light curve fit with \texttt{spotrod}. This corresponds to the joint fit of the orbital and starspot parameters. 
    \label{fig:corner_spot}}
\end{figure*}

 \begin{figure*}
	\centering
	\includegraphics[width=\textwidth]{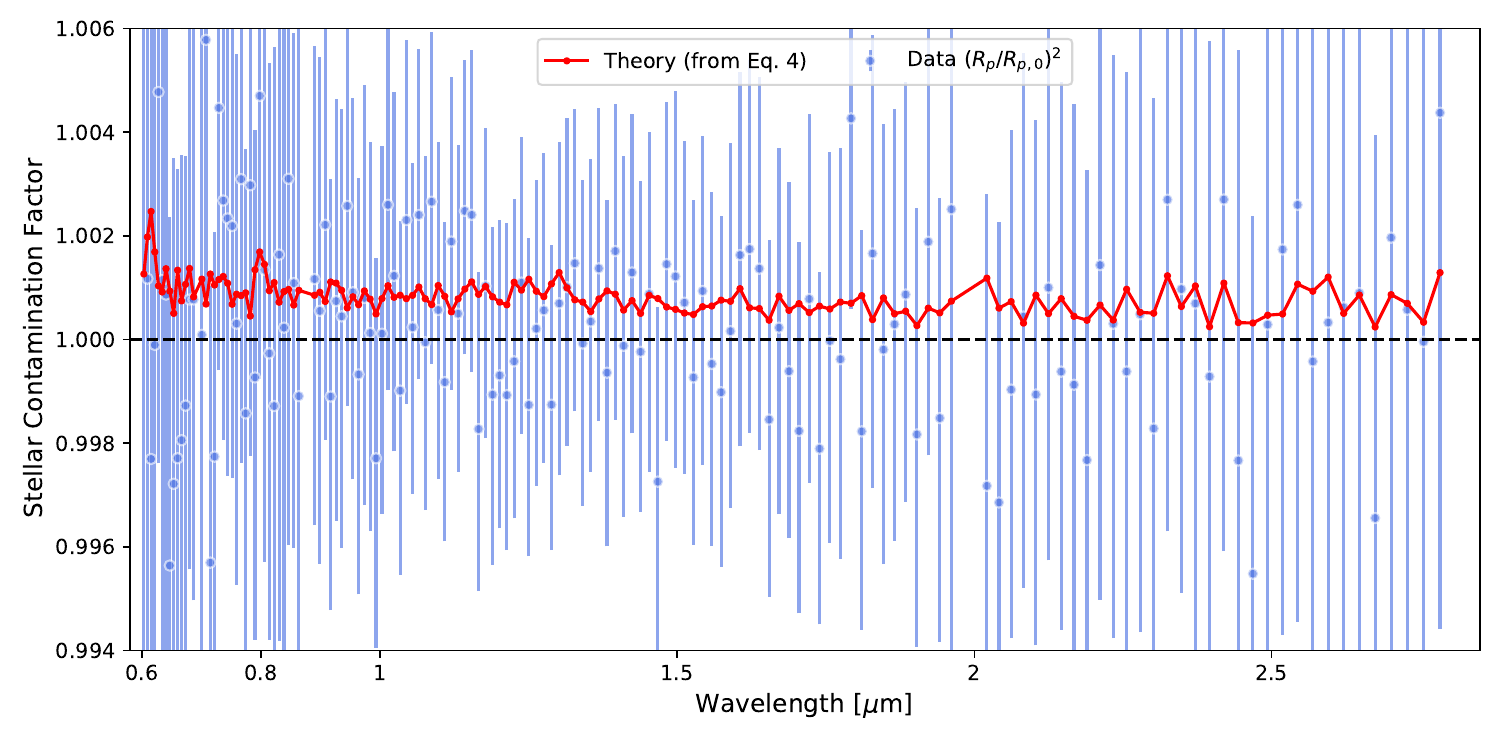}
    \caption{Stellar contamination on the transmission spectrum due to the occulted spot. The contamination factor (blue) represents the squared ratio of the planet's apparent radius \textit{R\textsubscript{p}} over the planet's true radius \textit{R\textsubscript{p,0}}. The former corresponds to a transit depth computed with a spot-crossing masked and the latter with a spot-crossing modelled. The theoretical factor (red) was derived from Eq. \ref{eq:stellar_contam_factor_one_het}.}
    \label{fig:occulted_spot}
\end{figure*}


\bsp	
\label{lastpage}
\end{document}